\documentclass[prd,aps,showpacs,nofootinbib,tightenlines]{revtex4}
\usepackage{mathrsfs}
\usepackage{amsmath}
\usepackage{amssymb}
\usepackage{epsfig}
\usepackage{graphicx}
\usepackage{booktabs}
\usepackage{multirow}
\usepackage{subfigure}
\usepackage[section]{placeins}
\usepackage{float}
\usepackage{slashed}
\begin{document}
\newcommand{\psl}{ p \hspace{-1.8truemm}/ }
\newcommand{\nsl}{ n \hspace{-2.2truemm}/ }
\newcommand{\vsl}{ v \hspace{-2.2truemm}/ }
\newcommand{\epsl}{\epsilon \hspace{-1.8truemm}/\,  }

\title{ Estimates of  exchange topological contributions and $CP$-violating observables in $\Lambda_b\rightarrow \Lambda \phi$  decay}
\author{Zhou Rui$^1$}\email[Corresponding  author: ]{jindui1127@126.com}
\author{Jia-Ming Li$^1$}
\author{Chao-Qi Zhang$^1$}
\affiliation{$^1$College of Sciences, North China University of Science and Technology,
Tangshan 063009,  China}
\date{\today}
\begin{abstract}
The penguin-dominated two-body weak decay of $\Lambda_b\rightarrow \Lambda \phi$ is studied based on the perturbative QCD approach.
In addition to the penguin emission  diagrams, the penguin exchange and $W$ exchange ones are also accounted for.
It is found that
the  penguin exchange  contribution is in fact important and comparable to the  penguin emission one,
while the $W$ exchange contribution is highly Cabibbo-Kobayashi-Maskawa (CKM)  suppressed.
The predicted branching ratio, $\mathcal{B}(\Lambda_b\rightarrow \Lambda \phi)=6.9^{+1.9+1.8}_{-2.0-1.6}\times10^{-6}$, is larger than the previous theoretical estimates
but in comparison with the data from Particle Data Group at the level of 1 standard deviation.
We also explore some pertinent decay asymmetry parameters that characterize the angular decay distributions.
The inclusion of the $W$ exchange contribution provides the nonzero weak phase difference,
consequently, allowing us to estimate the direct $CP$ violation and  true triple product asymmetries in the concerned process.
The numerical results  demonstrate that the direct $CP$ violation is at the level of a few percent,
and the true triple product asymmetries are also predicted to be tiny, of order $10^{-2}-10^{-4}$.
The observed small $CP$-violating observables have shown no significant deviations from zero.
Our predictions will be subject to stringent tests with precise data from LHCb in the future.

\end{abstract}



\maketitle

\section{Introduction}
With huge statistics of beauty hadrons were accumulated at the high energy and high intensity of the Large Hadron Collider (LHC),
several charmless $\Lambda_b$ decays of final states containing $\Lambda$ have been measured to be of order of $10^{-6}$
with good precision~\cite{LHCb:2013uqx,CDF:2011buy,LHCb:2019wwi,jhep090062015,plb759282,jhep050812016}.
After the evidence has been reported for the  $\Lambda_b\rightarrow \Lambda \eta$ decay~\cite{jhep090062015},
the first  vector mode $\Lambda_b\rightarrow \Lambda \phi$ was observed with a significance of 5.9 standard deviations by the LHCb Collaboration in 2016~\cite{plb759282}.
Some triple product asymmetries (TPAs) were measured to be consistent with zero and no $CP$ violation was found.
The current world average of its branching ratio given by Particle Data Group (PDG)~\cite{pdg2020} is $(9.8\pm2.6)\times10^{-6}$,
where multiple uncertainties are added in quadrature.
These measurements are  crucial for an in-depth  understanding of the strong dynamics  in the $b$-baryon decays.

Fueled by these observations, theoretical interests on the $\Lambda_b\rightarrow \Lambda \phi$ mode were increased recently.
In Ref.~\cite{prd99054020}, the authors studied the nonleptonic  two-body decays of
$\Lambda_b$ within the QCD factorization approach (QCDF) under the diquark hypothesis,
in which the two light spectator quarks in baryon are considered as a scalar of color antitriplet.
This approximation may work well for the  processes where the $W$ emission contribution is dominant.
However, for the processes where the diquark is broken during the transition, such as the  exchange topological contributions,
the plausibility of the diquark scenario may encounter serious challenges.
As the predicted branching ratio of the $\Lambda_b\rightarrow \Lambda \phi$ decay from~\cite{prd99054020} is an order of magnitude smaller than the experimental value,
the reasonability of the diquark hypothesis in this mode  needs to be further tested. 
The process $\Lambda_b\rightarrow \Lambda \phi$ was also addressed in~\cite{epjc76399,prd95093001} by using the generalized factorization approach (GFA),
in which the nonfactorizable contributions are parametrized in terms of the effective number of colors $N_c^{eff}$.
The obtained branching ratio with $N_c^{eff}$=2 was matched to the LHCb measurement,
which implies the sizeable nonfactorizable effects in the decay under consideration.
Very recently,
the angular analyses for $\Lambda_b\rightarrow \Lambda V$ with $V$ being a light vector meson has been derived in GFA~\cite{jhep111042021},
in which the $T$-violating observables were explored systematically.
Further information on this rich subject in the baryon sector may be found in Refs.~\cite{plb538309,prd63074001,plb614165,Leitner:2004dt,Leitner:2006nb,Leitner:2006sc,Gronau:2015gha,Durieux:2016nqr,Liu:2021rvt}.

Above theoretical studies are performed by the factorization ansatz,
within which the contributions from the factorizable  emission diagrams for the color-allowed decays can be estimated reliably.
Nonetheless, the nonfactorizable effects, especially from the $W$ and penguin exchange diagrams, still cannot be well explored.
As stated in~\cite{zpc55659,Mohanta:2000nk,prd562799,Uppal:1994pt}, contrary to the meson case, the $W$ exchange contribution plays a dramatic role in baryon decays
on account of the fact that
 it is neither helicity nor color suppressed since there may exist a scalar isosinglet diquark inside the baryon.
Experimentally
the observation of the $W$-exchange-only processes $\Lambda_c^+\rightarrow \Xi^0 K^+,\Delta^{++}K^-$~\cite{BESIII:2018cvs,pdg2020} with a surprisingly large branching ratios
indicates that $W$ exchange indeed plays an essential role in charmed baryon decays.
As of today,  many efforts have been made to estimate the $W$ exchange contributions to the baryon decays using phenomenological models such as
spectator quark model~\cite{zpc55659},
the pole model approach~\cite{Xu:1992vc,Zenczykowski:1993jm,Cheng:1993gf,Sharma:1998rd,Sharma:2017txj,Dhir:2018twm},
 the current-algebra approach~\cite{Sharma:1998rd,Uppal:1994pt,Cheng:2018hwl,Zou:2019kzq,Groote:2021pxt},
 final-state interaction  rescattering~\cite{Yu:2017zst,Jiang:2018oak},
the topological diagram  approach~\cite{Zhao:2018mov}, and constituent quark model~\cite{Gutsche:2018msz}.
All of these theoretical studies show that nonfactorizable $W$ exchange effects are generally important in the weak decays of singly and doubly charm baryons.
Recently,
it was also observed that the $W$ exchange contribution is in fact not negligible in the charmful two-body baryonic $B$ decays~\cite{Hsiao:2019wyd}.
This begs the question of whether $W$ exchange effects are important in bottom baryon decays,
which have  received less attention in the literature except those in Refs.~\cite{Ivanov:1997ra,Ivanov:1997hi,221010017}.

In the bottom baryon sector,
the significant $W$ exchange contribution in the exclusive nonleptonic bottom to charm baryon decays was observed to be
as important as the nonfactorizable $W$ emission one based on the relativistic three-quark model~\cite{Ivanov:1997ra,Ivanov:1997hi}.
In the decay $\Lambda_b\rightarrow \Lambda_c \pi$, the total contribution of the nonfactorizable diagrams can amount to $30\%$ of the factorizable contribution in amplitude.
Nevertheless, the discussion in~\cite{Ivanov:1997ra,Ivanov:1997hi} is only limited to the Cabibbo-favored decays mediated
by the tree type transition $b\rightarrow c\bar u d$ with a light pseudoscalar meson in the final state.
Very recently, the $W$ exchange contribution has been studied in some hadronic decays of bottom baryon~\cite{221010017},
where the  initial and final state baryons belong to different isospin representations and any factorizable amplitude is forbidden.
It will be of great interest to  examine how substantial  these nonfactorizable effects are in the penguin-dominated processes.
On the other hand,
it is well known that the $T$-odd observables vanish in the penguin-dominated decays  induced by a single weak $b\rightarrow s$ transition.
However, after considering $W$ exchange tree amplitudes, they could interfere with the penguin amplitudes,
and produce the nonzero values of true $T$-violating observables,
which could be measured in the experiments.

The perturbative QCD (PQCD) approach has been developed and successfully applied to deal with weak decays of $\Lambda_b$ baryon
\cite{prd59094014,prd61114002,cjp39328,prd74034026,prd65074030,prd80034011,220204804},
in which various contributions, such as emission and exchange ones,  can be evaluated in a self-consistent manner.
We previously investigated  the color-allowed $\Lambda_b\rightarrow \Lambda_c (\pi, K)$ decays~\cite{prd105073005}
and the color-suppressed $\Lambda_b\rightarrow \Lambda (J/\psi,\psi(2S))$ decays~\cite{220604501}
within this approach and obtained satisfactory results.
In this work, we will analyze the two-body $\Lambda_b\rightarrow\Lambda \phi$ decay in the PQCD approach to the leading order in the strong coupling $\alpha_s$ expansion.
The decay under consideration
is the penguin-dominated mode  induced by the neutral-current $b\rightarrow s $ transition,
which also receive tree diagram contributions from the $W$ exchange diagrams via $bu\rightarrow su$ transition.
The decay amplitudes include the contributions from the penguin emission, penguin exchange, as well as $W$ exchange diagrams.
The last two contributions were not considered in previous studies.
By explicitly calculating this process,
we shall demonstrate that the penguin exchange diagrams in fact give contribution of the same order as that from penguin emission ones,
while the $W$ exchange contribution is highly CKM suppressed.
Aside from the decay branching ratio, many asymmetries derived from the angular distribution are also predicted and compared with other theoretical results and experiments.
In particular, we give the theoretical predictions on the TPAs in $\Lambda_b\rightarrow \Lambda \phi$ decay for the first time,
which could be checked by future experiments.

The rest of the paper is organized as follows.
After presenting the effective hamiltonian,  kinematics, and the light-cone distribution amplitudes (LCDAs) in Sec.~\ref{sec:framework},
we give a general PQCD formalism for $\Lambda_b\rightarrow \Lambda \phi$,
which can be applied to other $\Lambda_b\rightarrow \Lambda V$ decays.
The numerical results for the  invariant and helicity amplitudes, decay branching
ratio,  various asymmetries, and $CP$-violating observables are presented in  Sec.~\ref{sec:results}.
Finally, Sec.~\ref{sec:sum} will be the conclusion of this work.
Some details of the factorization formulas are displayed in Appendix~\ref{sec:for}.

\section{Theoretical framework}\label{sec:framework}
\begin{figure}[!htbh]
	\begin{center}
		\vspace{0.01cm} \centerline{\epsfxsize=15cm \epsffile{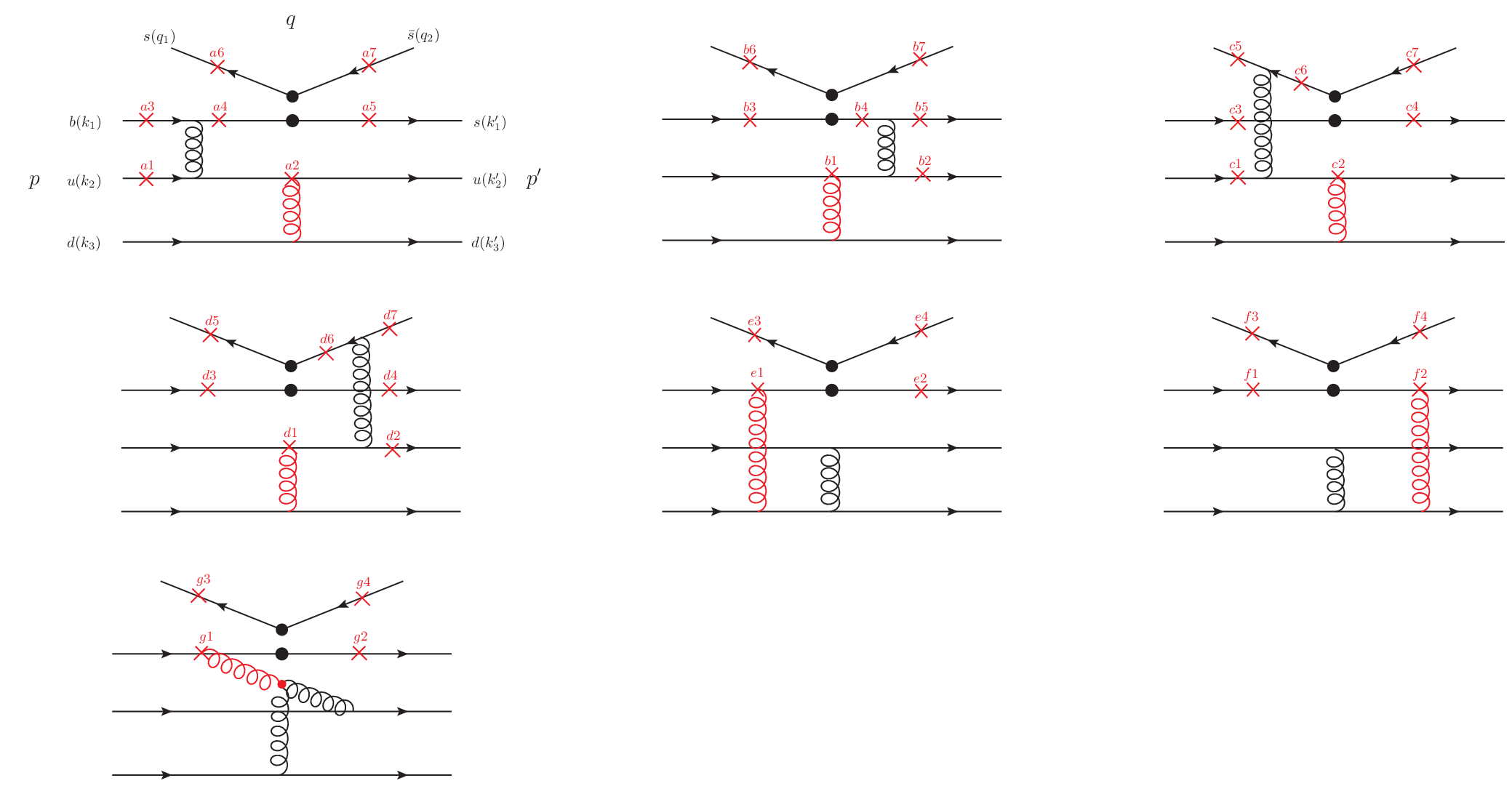}}
		\setlength{\abovecaptionskip}{1.0cm}
		\caption{Penguin emission ($P$) diagrams for the $\Lambda_b\rightarrow \Lambda \phi$ decay to the lowest order in the PQCD approach,
           where the solid black blob represents the vertex of the effective weak interaction.
           The crosses on the quark lines, indicated by $ij$ with $i=a-f$ and $j=1-7$, denote the possible ways in which the quark is connected to the spectator $d$ quark via a hard gluon.
          Those diagrams with exchanging $u$ and $d$ quarks in the $\Lambda_{(b)}$ baryons simultaneously, giving the identical contribution, are not displayed.}
		\label{fig:FeynmanT}
	\end{center}
\end{figure}

\begin{figure}[!htbh]
	\begin{center}
		\vspace{0.01cm} \centerline{\epsfxsize=15cm \epsffile{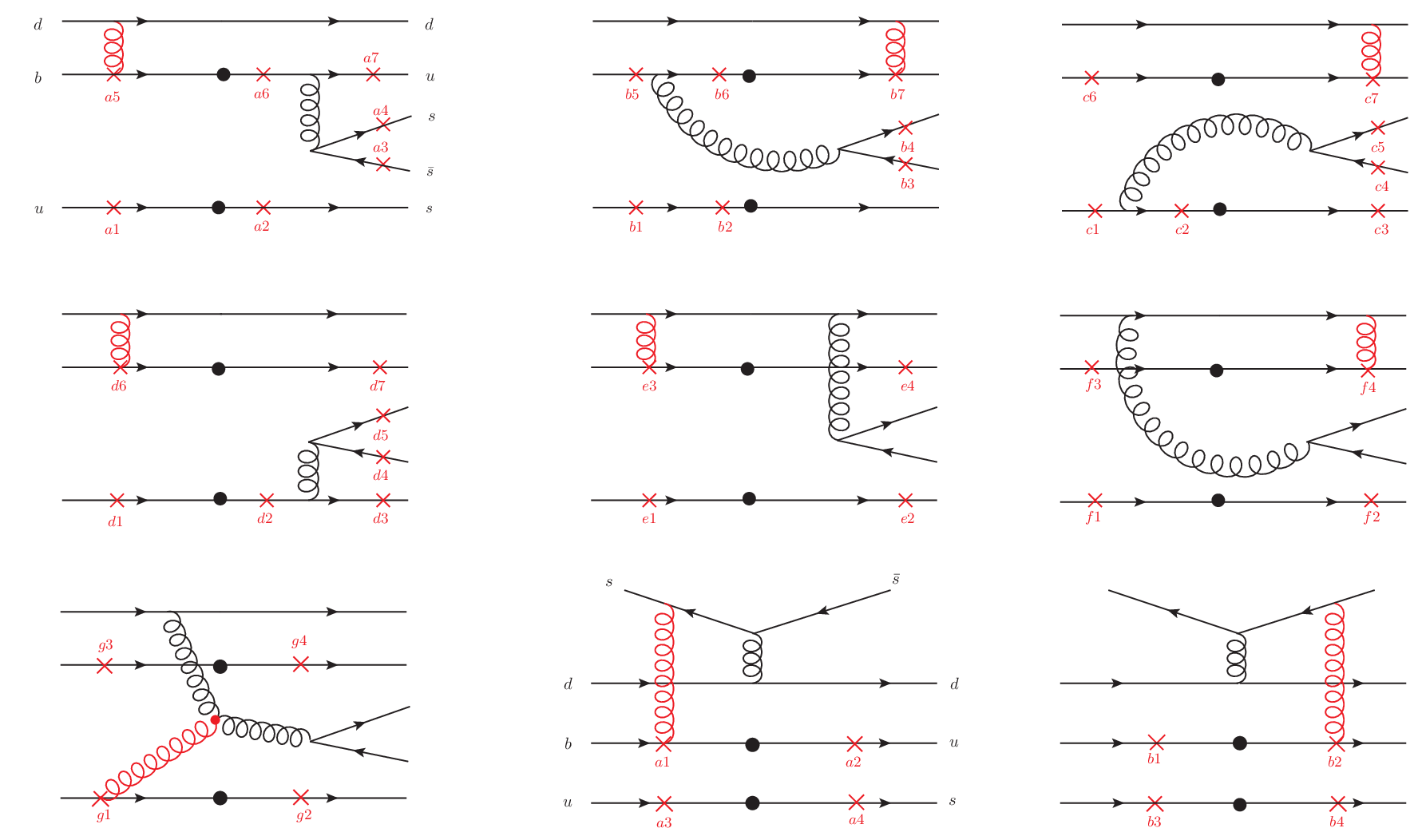}}
		\setlength{\abovecaptionskip}{1.0cm}
		\caption{$W$ exchange diagrams for the decay $\Lambda_b\rightarrow \Lambda \phi$.
The first two rows  are called $E$-type diagrams marked by $E_{ij}$ with $i=a-f$ and $j=1-7$.
The first diagram in the third row is the three-gluon $E$-type marked by $E_{gj}$ with $j=1-4$, while
the last two diagrams are classified as bow tie type $W$ exchange diagrams marked by $B_{ij}$ with $i=a,b$ and $j=1-4$.}
		\label{fig:FeynmanE}
	\end{center}
\end{figure}

\begin{figure}[!htbh]
	\begin{center}
		\vspace{0.01cm} \centerline{\epsfxsize=15cm \epsffile{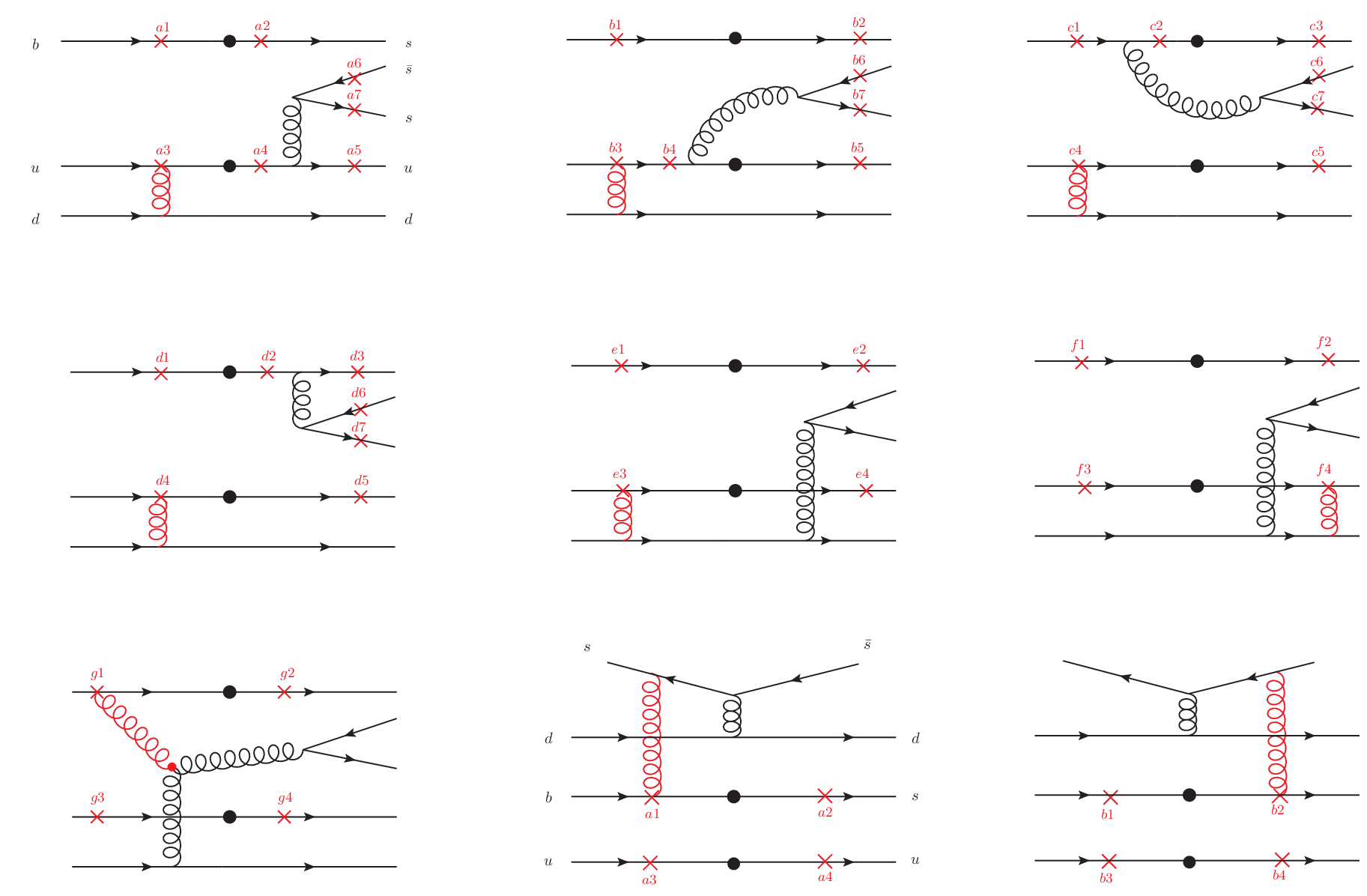}}
		\setlength{\abovecaptionskip}{1.0cm}
		\caption{Penguin exchange diagrams for the decay $\Lambda_b\rightarrow \Lambda \phi$,
       which are similar to Fig.~\ref{fig:FeynmanE} but with the penguin operators inserting.}
		\label{fig:FeynmanP}
	\end{center}
\end{figure}

\subsection{Hamiltonian and kinematics}\label{sec:Kinematics}
In the standard model (SM), $\Lambda_b\rightarrow \Lambda \phi$ involving $b\rightarrow s $ transition
is governed by the following effective Hamiltonian~\cite{Buchalla:1995vs}
\begin{eqnarray}
\mathcal{H}_{eff}&=&\frac{G_F}{\sqrt{2}} \left(V_{ub}V^*_{us}[C_1(\mu)O_1(\mu)+C_2(\mu)O_2(\mu)]-\sum_{k=3}^{10}V_{tb}V^*_{ts}C_k(\mu)O_k(\mu)\right)+H.c.,
\end{eqnarray}
with the Fermi coupling constant $G_F$.
$V_{ij}$  are the related  Cabibbo-Kobayashi-Maskawa (CKM) matrix elements.
$C_i(\mu)$ are the Wilson coefficients evaluated at the renormalization scale $\mu$.
The effective four-quark  operators $O_i$ containing quark and gluon fields are given by
\begin{eqnarray}
O_1&=& \bar{u}_\alpha \gamma_\mu(1-\gamma_5) b_\beta  \otimes \bar{s}_\beta  \gamma^\mu(1-\gamma_5) u_\alpha, \nonumber\\
O_2&=& \bar{u}_\alpha \gamma_\mu(1-\gamma_5) b_\alpha \otimes \bar{s}_\beta  \gamma^\mu(1-\gamma_5) u_\beta,  \nonumber\\
O_3&=& \bar{s}_\beta  \gamma_\mu(1-\gamma_5) b_\beta  \otimes \sum_{q'} \bar{q}'_\alpha \gamma^\mu(1-\gamma_5) q'_\alpha, \nonumber\\
O_4&=& \bar{s}_\beta  \gamma_\mu(1-\gamma_5) b_\alpha \otimes \sum_{q'} \bar{q}'_\alpha \gamma^\mu(1-\gamma_5) q'_\beta,  \nonumber\\
O_5&=& \bar{s}_\beta  \gamma_\mu(1-\gamma_5) b_\beta  \otimes \sum_{q'} \bar{q}'_\alpha \gamma^\mu(1+\gamma_5) q'_\alpha, \nonumber\\
O_6&=& \bar{s}_\beta  \gamma_\mu(1-\gamma_5) b_\alpha \otimes \sum_{q'} \bar{q}'_\alpha \gamma^\mu(1+\gamma_5) q'_\beta,  \nonumber\\
O_7&=&   \frac{3}{2}\bar{s}_\beta \gamma_\mu(1-\gamma_5) b_\beta  \otimes \sum_{q'} e_{q'}\bar{q}'_\alpha \gamma^\mu(1+\gamma_5) q'_\alpha, \nonumber\\
O_8&=&   \frac{3}{2}\bar{s}_\beta \gamma_\mu(1-\gamma_5) b_\alpha \otimes \sum_{q'} e_{q'}\bar{q}'_\alpha \gamma^\mu(1+\gamma_5) q'_\beta,  \nonumber\\
O_9&=&   \frac{3}{2}\bar{s}_\beta \gamma_\mu(1-\gamma_5) b_\beta  \otimes \sum_{q'} e_{q'}\bar{q}'_\alpha \gamma^\mu(1-\gamma_5) q'_\alpha, \nonumber\\
O_{10}&=&\frac{3}{2}\bar{s}_\beta \gamma_\mu(1-\gamma_5) b_\alpha \otimes \sum_{q'} e_{q'}\bar{q}'_\alpha \gamma^\mu(1-\gamma_5) q'_\beta,
\end{eqnarray}
where $O_{1,2}$ are the current-current operators arising from $W$ boson exchange ($\alpha,\beta$  denote colors),
$O_{3,4,5,6}$ and $O_{7,8,9,10}$ are the QCD and the electroweak penguin operators, respectively.
$e_{q'}$ is the electric charges of the quark $q'$ in units of $|e|$.
 The sum over $q'$ runs over the quark fields active at the $b$ quark mass
scale, i.e., $q'=u,d,s,c,b$.

According to the topological  classification  of weak interactions,
the concerned decay amplitude
receives  contributions from
the penguin emission, $W$ exchange,
and penguin exchange diagrams as shown in Figs.~\ref{fig:FeynmanT}, \ref{fig:FeynmanE}, and \ref{fig:FeynmanP}, respectively.
The emission diagrams shown in Fig.~\ref{fig:FeynmanT} can only contribute to the decay by inserting the penguin operators,
 we thus call them the penguin emission diagrams marked by $P$.
Figure~\ref{fig:FeynmanE} is manifested in the $W$ exchange processes via $bu\rightarrow su$ transition,
which can be classified into two types labeled by $E$ and $B$ (the last two diagrams), respectively.
The former represents that the two quarks produced by the weak interaction are shared by the final state baryon and meson,
while the latter denotes that both the two quarks flow into the $\Lambda$ baryon.
By exchanging the $b$ and $u$ quarks in the $\Lambda_b$ baryon from the $W$ exchange diagrams,
one can obtain two corresponding penguin exchange diagrams denoted by $PE$ and $PB$ as exhibited in Fig.~\ref{fig:FeynmanP}.
We do not show those diagrams by exchanging $u$ and $d$ quarks in the initial and final states baryons from the previous diagrams because their amplitudes are equivalent under this interchange.
In total, there are 136 Feynman diagrams,
each denoted by $R_{ij}$ with $R=P, PE, PB, E, B$
and the subscript $ij$ representing  possible ways of exchanging two hard gluons.
It is worth to underline that in the $B$ and $PB$ type diagrams,
the $s \bar s$ pair must attach two gluons to form a color singlet $\phi$ meson.
Since the fermion flows of the diagrams $PE(PB)$ and $E(B)$ type can be converted into each other via the Fierz transformation,
we insert the tree operators into $E$ and $B$ type diagrams and the penguin
operators into  $PE$ and $PB$ ones in the following analysis to avoid counting these contributions repetitively.

We shall work in the rest frame of the baryon $\Lambda_b$  with the baryon $\Lambda$  moving in the dominant positive direction on the light cone
\begin{eqnarray}\label{eq:pq}
p=\frac{M}{\sqrt{2}}(1,1,\textbf{0}_{T}), \quad p'=\frac{M}{\sqrt{2}}(f^+,f^-,\textbf{0}_{T}), \quad
q=\frac{M}{\sqrt{2}}\left(1-f^+,1-f^-,\textbf{0}_{T}\right),
\end{eqnarray}
with $M$ being the $\Lambda_b$ baryon mass.
The factors $f^{\pm}$ can be derived from the on-shell conditions $p'^2=m_\Lambda^2$ and $q^2=m_\phi^2$ for the final-state hadrons, which yield
\begin{eqnarray}
f^\pm=\frac{1}{2}\left(1-r_\phi^2+r_{\Lambda}^2 \pm \sqrt{(1-r_\phi^2+r_{\Lambda }^2)^2-4r_{\Lambda}^2}\right),
\end{eqnarray}
with the mass ratios $r_{\Lambda,\phi}=m_{\Lambda,\phi}/M$.
At the end of the derivation of the factorization formulas, the terms $r^2_{\Lambda,\phi}\sim 0.04$ in the above kinematic variables will be neglected.
For the vector meson,
the longitudinal and transverse polarization vectors ($\epsilon_{L,T}$) can be determined by
the normalization and orthogonality conditions as
\begin{eqnarray}
\epsilon_L=\frac{1}{\sqrt{2}r_\phi}\left(f^+-1,1-f^-,\textbf{0}_{T}\right),
\quad \epsilon_T=\left(0,0,\textbf{1}_{T}\right).
\end{eqnarray}

The momenta of eight valence quarks in the initial and final states, whose notations are displayed in Fig.~\ref{fig:FeynmanT}, are parametrized as
\begin{eqnarray}
k_1&=&\left(\frac{M}{\sqrt{2}},\frac{M}{\sqrt{2}}x_1,\textbf{k}_{1T}\right),\quad
k_2=\left(0,\frac{M}{\sqrt{2}}x_2,\textbf{k}_{2T}\right),\quad
k_3=\left(0,\frac{M}{\sqrt{2}}x_3,\textbf{k}_{3T}\right),\nonumber\\
k_1'&=&\left(\frac{M}{\sqrt{2}}f^+x_1',0,\textbf{k}'_{1T}\right),\quad
k_2'=\left(\frac{M}{\sqrt{2}}f^+x_2',0,\textbf{k}'_{2T}\right),\quad
k_3'=\left(\frac{M}{\sqrt{2}}f^+x_3',0,\textbf{k}'_{3T}\right),\nonumber\\
q_1&=&\left(\frac{M}{\sqrt{2}}y(1-f^+),\frac{M}{\sqrt{2}}y(1-f^-),\textbf{q}_{T}\right),\nonumber\\
q_2&=&\left(\frac{M}{\sqrt{2}}(1-y)(1-f^+),\frac{M}{\sqrt{2}}(1-y)(1-f^-),-\textbf{q}_{T}\right),
\end{eqnarray}
where  $x^{(')}_{1,2,3}$, and $y$ are the parton longitudinal momentum fractions and
$\textbf{k}^{(')}_{1T,2T,3T}$, and $\textbf{q}_T$ are the corresponding transverse momenta.
The momentum conservation implies the relations
\begin{eqnarray}
\sum_{l=1}^3x^{(')}_l=1,\quad \sum_{l=1}^3\textbf{k}^{(')}_{lT}=0.
\end{eqnarray}

\subsection{Light-cone distribution amplitudes}\label{sec:LCDAs}
In the heavy quark limit, the $\Lambda_b$ baryon LCDAs can be defined as matrix elements of non-local light-ray operators~\cite{plb665197,jhep112013191,epjc732302,plb738334,jhep022016179,Ali:2012zza}.
It is more convenient to use the form in the momentum-space for practical applications,
whose  Lorentz structures up to twist-4 accuracy can be written as~\cite{jhep112013191}
\begin{eqnarray}
(\Psi_{\Lambda_b})_{\alpha\beta\gamma}(x_i,\mu)=\frac{1}{8N_c}
\{f_{\Lambda_b}^{(1)}(\mu)[M_1(x_2,x_3)\gamma_5C^T]_{\gamma\beta}
+f_{\Lambda_b}^{(2)}(\mu)[M_2(x_2,x_3)\gamma_5C^T]_{\gamma\beta}\}[\Lambda_b(p)]_\alpha,
\end{eqnarray}
where $\alpha,\beta,\gamma$ are the spinor indices.
$\Lambda_b(p)$ is the Dirac spinor.
$N_c$ is the number of colors.
$C^T$ denotes the charge conjugation matrix under transpose transform.
The normalization constants $f_{\Lambda_b}^{(1)}\approx f_{\Lambda_b}^{(2)}\equiv f_{\Lambda_b}=0.021\pm0.004$ GeV$^3$~\cite{220204804},
which is close to the value from the leading-order sum rule calculation~\cite{Groote:1997yr}.
The chiral-even ($M_{1}$) and odd ($M_{2}$) projectors  read
\begin{eqnarray}
M_1(x_2,x_3)&=&\frac{\slashed{v}\slashed{n}}{4}\Psi_3^{+-}(x_2,x_3)
+\frac{\slashed{n}\slashed{v}}{4}\Psi_3^{-+}(x_2,x_3), \nonumber\\
M_2(x_2,x_3)&=&\frac{\slashed{n}}{\sqrt{2}}\Psi_2(x_2,x_3)
+\frac{\slashed{v}}{\sqrt{2}}\Psi_4(x_2,x_3),
\end{eqnarray}
respectively,
where two light-cone vectors $n=(1,0,\textbf{0}_T)$ and $v=(0,1,\textbf{0}_T)$ satisfy $n\cdot v=1$.
Note that the momentum of the $\Lambda$  baryon is along the $n$ direction in the massless limit.
For the shape of the various twist, we use the simple exponential model~\cite{jhep112013191}
\begin{eqnarray}
\Psi_2(x_2,x_3)&=&     x_2x_3\frac{M^4}{\omega_0^4}e^{-\frac{(x_2+x_3)M}{\omega_0}},\nonumber\\
\Psi_3^{+-}(x_2,x_3)&=&2x_2\frac{M^3}  {\omega_0^3}e^{-\frac{(x_2+x_3)M}{\omega_0}},\nonumber\\
\Psi_3^{-+}(x_2,x_3)&=&2x_3\frac{M^3}  {\omega_0^3}e^{-\frac{(x_2+x_3)M}{\omega_0}},\nonumber\\
\Psi_4(x_2,x_3)&=&     \frac{ M^2}     {\omega_0^2}e^{-\frac{(x_2+x_3)M}{\omega_0}},
\end{eqnarray}
with $\omega_0=0.4$ GeV.

For the LCDAs of $\Lambda$ baryon, we would like to adopt the Chernyak-Ogloblin-Zhitnitsky (COZ) model proposed in Ref.~\cite{zpc42569}.
This choose is supported by the previous PQCD calculation on the form factors of $\Lambda_b\rightarrow \Lambda$ transition as discussed in~\cite{220604501}. 
Other available models, such as QCD sum rules and lattice QCD (LQCD),
one refers to Refs.~\cite{Wang:2008sm,Liu:2014uha,Liu:2008yg,jhep020702016,prd89094511,epja55116} for details.
The nonlocal matrix element associated with $\Lambda$ baryon at leading twist is given by~\cite{220604501}
\begin{eqnarray}\label{eq:wave}
(\Psi_{\Lambda})_{\alpha\beta\gamma}(k_i',\mu)
&=&\frac{1}{8\sqrt{2}N_c}
\{(\slashed{p}'C)_{\beta\gamma}[\gamma_5\Lambda({p}')]_\alpha\Phi^V(k_i',\mu)
+(\slashed{p}'\gamma_5C)_{\beta\gamma}[\Lambda({p}')]_\alpha\Phi^A(k_i',\mu) \nonumber\\
&&+(i\sigma_{\mu\nu}{p'}^\nu C)_{\beta\gamma}[\gamma^\mu\gamma_5\Lambda({p}')]_\alpha\Phi^T(k_i',\mu)\},
\end{eqnarray}
with $\sigma_{\mu\nu}=i[\gamma_\mu,\gamma_\nu]/2$, and $\Lambda(p')$ is the $\Lambda$ baryon spinor.
$\Phi^{V}$ and $\Phi^{T}$ are antisymmetric under permutation of two light quarks,
while $\Phi^A$ is symmetric under the same operation.
Their explicit forms at the scale $\mu=1$ GeV in the COZ model are given as~\cite{zpc42569}
\begin{eqnarray}\label{eq:vat}
\Phi^V(x_1,x_2,x_3)&=&42f_{\Lambda}\phi_{asy}[0.18(x_2^2-x_3^2)-0.1(x_2-x_3)],\nonumber\\
\Phi^A(x_1,x_2,x_3)&=&-42f_{\Lambda}\phi_{asy}[0.26(x_3^2+x_2^2)+0.34x_1^2-0.56x_2x_3-0.24x_1(x_2+x_3)],\nonumber\\
\Phi^T(x_1,x_2,x_3)&=&42f_{\Lambda}^T\phi_{asy}[1.2(x_3^2-x_2^2)+1.4(x_2-x_3)],
\end{eqnarray}
with $\phi_{asy}(x_1,x_2,x_3)=120x_1x_2x_3$ being the asymptotic form in the limit of $\mu\rightarrow \infty$.
$\Phi^A$ and $\Phi^{T}$  satisfy the normalizations~\cite{zpc42569}
\begin{eqnarray}
\int_0^1\Phi^A dx_1dx_2dx_3\delta(1-x_1-x_2-x_3)=-f_{\Lambda}, \quad \int_0^1\Phi^Tx_2dx_1dx_2dx_3\delta(1-x_1-x_2-x_3)=f_{\Lambda}^T,
\end{eqnarray}
respectively, where the two normalization constants are set to be $f_{\Lambda}=10f^T_{\Lambda}=6.3\times10^{-3}$ GeV$^2$~\cite{zpc42569}.

For a light vector meson $V$, the light-cone distribution amplitudes for longitudinal ($L$) and transverse ($T$) polarizations can   be written as~\cite{Ball:1998sk,Ball:1998ff,Ball:2007rt}
\begin{eqnarray}
\Phi_V^L(y)&=&\frac{1}{\sqrt{2N_c}}[m_V\rlap{/}{\epsilon_L}\phi_V(y)
+\rlap{/}{\epsilon_L}\rlap{/}{q}\phi_V^t(y)+m_V\phi_V^s(y)],\nonumber\\
\Phi_V^T(y)&=&\frac{1}{\sqrt{2N_c}}[m_V\rlap{/}{\epsilon_T}\phi_V^v(y)
+\rlap{/}{\epsilon_T}\rlap{/}{q}\phi_V^T(y)+im_V\epsilon^{\mu\nu\rho\sigma}
\gamma_5\gamma_{\mu}\epsilon_{T\nu}v_{\rho}n_{\sigma}\phi_V^a(y)],
\end{eqnarray}
respectively.
Here  $\epsilon^{\mu\nu\rho\sigma}$ is the totally antisymmetric unit
Levi-Civita tensor with the convention $\epsilon^{0123}=1$.
The twist-2 LCDAs are given by
\begin{eqnarray}\label{eq:twist2}
\phi_V(y)&=&\frac{f_V}{\sqrt{2N_c}}3y(1-y)[1+a_{1V}^{\parallel}3(2y-1)+a_{2V}^{\parallel}3(5(2y-1)^2-1)/2],\nonumber\\
\phi_V^T(y)&=&\frac{f_V^T}{\sqrt{2N_c}}3y(1-y)[1+a_{1V}^{\perp}3(2y-1)+a_{2V}^{\perp}3(5(2y-1)^2-1)/2],
\end{eqnarray}
 and those of twist-3 ones for the asymptotic form are
\begin{eqnarray}
\phi_V^t(y)&=&\frac{3f_V^T}{2\sqrt{2N_c}}(2y-1)^2,\quad \phi_V^s(y)=-\frac{3f_V^T}{2\sqrt{2N_c}}(2y-1),\nonumber\\
\phi_V^v(y)&=&\frac{3f_V}{8\sqrt{2N_c}}[1+(2y-1)^2],\quad \phi_V^a(y)=-\frac{3f_V}{4\sqrt{2N_c}}(2y-1).
\end{eqnarray}
The values of the Gegenbauer moments and decay constants for $\phi$ meson are taken as~\cite{Ball:2007rt,Rui:2017fje}
\begin{eqnarray}
a_{1}^{\parallel}=a_{1}^{\perp}=0,\quad a_{2}^{\parallel}=0.18\pm 0.08, \quad a_{2}^{\perp}=0.14\pm0.07,\quad f_\phi=(215\pm5)~\text{MeV},\quad f_\phi^T=(186\pm9)~\text{MeV}.
\end{eqnarray}

\subsection{Invariant amplitudes and helicity amplitudes}\label{sec:amplitudes}
The decay amplitude for $\frac{1}{2}^+\rightarrow \frac{1}{2}^+ + 1^-$ type can be expanded with the Dirac spinors and polarization vector as~\cite{zpc55659,prd562799},
\begin{eqnarray}\label{eq:kq}
\mathcal{M}^L&=&\bar {\Lambda} (p')\epsilon^{\mu*}_{L}[A_1^L\gamma_{\mu}\gamma_5+A_2^L\frac{p'_{\mu}}{M}\gamma_5+B_1^L\gamma_\mu+B_2^L\frac{p'_{\mu}}{M}]\Lambda_b(p),
\nonumber\\
\mathcal{M}^T&=&\bar {\Lambda} (p')\epsilon^{\mu*}_{T}[A_1^T\gamma_{\mu}\gamma_5+B_1^T\gamma_\mu]\Lambda_b(p),
\end{eqnarray}
where we split the amplitude into the longitudinal and transverse pieces according to the vector meson polarizations~\cite{220604501}.
Above we have included explicit factors of $M$ so that $A(B)_2^L$ have the same dimensions as $A(B)_1^L$.
The terms $A$ and $B$ denote the parity-violating  and parity-conserving amplitudes, respectively.
Their general factorization formula in PQCD can symbolically  be written as
\begin{eqnarray}\label{eq:amp}
A(B)=\frac{f_{\Lambda_b}\pi^2 G_F}{18\sqrt{3}}\sum_{R_{ij}}
\int[\mathcal{D}x][\mathcal{D}b]_{R_{ij}}
\alpha_s^2(t_{R_{ij}})\Omega_{R_{ij}}(b,b',b_q)e^{-S_{R_{ij}}}\sum_{\sigma=LL,LR,SP}a_{R_{ij}}^{\sigma}H^{\sigma}_{R_{ij}}(x,x',y),
\end{eqnarray}
where the summation extends over all possible diagrams $R_{ij}$.
 $a_{R_{ij}}^{\sigma}$ denotes the product of the CKM matrix elements and the Wilson coefficients,
and the labels $\sigma=LL$,  $LR$, and $SP$ refer to the contributions from $(V-A)(V-A)$, $(V-A)(V+A)$, and $(S-P)(S+P)$ operators, respectively.
Here $b$, $b'$ and $b_q$ are the conjugate variables to $k_T$, $k'_T$ and $q_T$, respectively.
$H^{\sigma}_{R_{ij}}$ is the numerator of the hard amplitude depending on the spin structure of final state.
$\Omega_{R_{ij}}$ is the Fourier transformation of the denominator of the hard amplitude from the $k_T$ space to its conjugate $b$ space.
The integration measure of the momentum fractions are defined as
\begin{eqnarray}
[\mathcal{D}x]=[dx_1dx_2dx_3\delta(1-x_1-x_2-x_3)][dx'_1dx'_2dx'_3\delta(1-x'_1-x'_2-x'_3)]dy,
\end{eqnarray}
where the $\delta$ functions enforce momentum conservation.
The hard scale $t$ for each diagram is chosen as the maximal virtuality of internal particles
including the factorization scales in a hard amplitude:
\begin{eqnarray}
t_{R_{ij}}=\max(\sqrt{|t_A|},\sqrt{|t_B|},\sqrt{|t_C|},\sqrt{|t_D|},w,w',w_q),
\end{eqnarray}
where the expressions of $t_{A,B,C,D}$ will be listed in Appendix~\ref{sec:for}.
The factorization scales $w$, $w'$, and $w_q$ are defined by
 \begin{eqnarray}
w^{(')}=\min(\frac{1}{b^{(')}_1},\frac{1}{b^{(')}_2},\frac{1}{b^{(')}_3}),\quad w_q=\frac{1}{b_q},
\end{eqnarray}
 with the variables
 \begin{eqnarray}
b^{(')}_1=|b^{(')}_2-b^{(')}_3|,
\end{eqnarray}
and the other $b^{(')}_l$ defined by permutation.
The explicit forms of the Sudakov factors $S_{R_{ij}}$ can be found in~\cite{220604501}.
Those quantities associated with specific diagram, such as $H_{R_{ij}}$, $a_{R_{ij}}$, $[\mathcal{D}b]_{R_{ij}}$, and $t_{R_{ij}}$, are collected in Appendix~\ref{sec:for}.

It is convenient to  apply the helicity amplitudes $H_{\lambda_\Lambda\lambda_\phi}$ for expressing various observable quantities in the decays,
where $\lambda_{\Lambda}$ and  $\lambda_{\phi}$ are the respective helicities of $\Lambda$ and $\phi$ with the possible values $\lambda_{\Lambda}=\pm 1/2$ and $\lambda_{\phi}=0,\pm1$.
The helicity of the $\Lambda_b$ baryon $\lambda_{\Lambda_b}$ is related by $\lambda_{\Lambda_b}=\lambda_{\Lambda}-\lambda_{\phi}$~\cite{zpc55659}.
Angular momentum conservation allows four independent helicity amplitudes to contribute,
including two transverse polarizations $H_{\pm\frac{1}{2}\pm1}$ and two longitudinal ones $H_{\pm\frac{1}{2}0}$,
which can be expressed in terms of the invariant amplitudes $A$ and $B$ as~\cite{zpc55659,prd562799}
\begin{eqnarray}\label{eq:helicity}
H_{\frac{1}{2}1}&=&-(\sqrt{Q_+}A_1^T+\sqrt{Q_-}B_1^T),\nonumber\\
H_{-\frac{1}{2}-1}&=&\sqrt{Q_+}A_1^T-\sqrt{Q_-}B_1^T,\nonumber\\
H_{\frac{1}{2}0}&=& \frac{1}{\sqrt{2}m_\phi}[\sqrt{Q_+}(M-m_\Lambda)A_1^L-\sqrt{Q_-}P_cA_2^L+\sqrt{Q_-}(M+m_\Lambda)B_1^L+\sqrt{Q_+}P_cB_2^L],\nonumber\\
H_{-\frac{1}{2}0}&=& \frac{1}{\sqrt{2}m_\phi}[-\sqrt{Q_+}(M-m_\Lambda)A_1^L+\sqrt{Q_-}P_cA_2^L+\sqrt{Q_-}(M+m_\Lambda)B_1^L+\sqrt{Q_+}P_cB_2^L],
\end{eqnarray}
with $Q_{\pm}=(M\pm m_\Lambda)^2-m_\phi^2$. $P_c=\sqrt{Q_+Q_-}/(2M)$ is the $\Lambda$ momentum in the center of $\Lambda_b$ mass frame.

The two-body decay branching ratio reads
\begin{eqnarray}\label{eq:two}
\mathcal{B}=\frac{P_c\tau_{\Lambda_b}}{8\pi M^2}H_N,
\end{eqnarray}
where the sum of the magnitude squared of the helicity amplitudes $H_N$ is denoted by
\begin{eqnarray}\label{eq:hn}
H_N=|H_{\frac{1}{2}1}|^2+|H_{-\frac{1}{2}-1}|^2+|H_{\frac{1}{2}0}|^2+|H_{-\frac{1}{2}0}|^2.
\end{eqnarray}

\subsection{Angular distributions and triple product asymmetries}\label{sec:tpas}
\begin{figure}[!htbh]
	\begin{center}
		\vspace{0.01cm} \centerline{\epsfxsize=15cm \epsffile{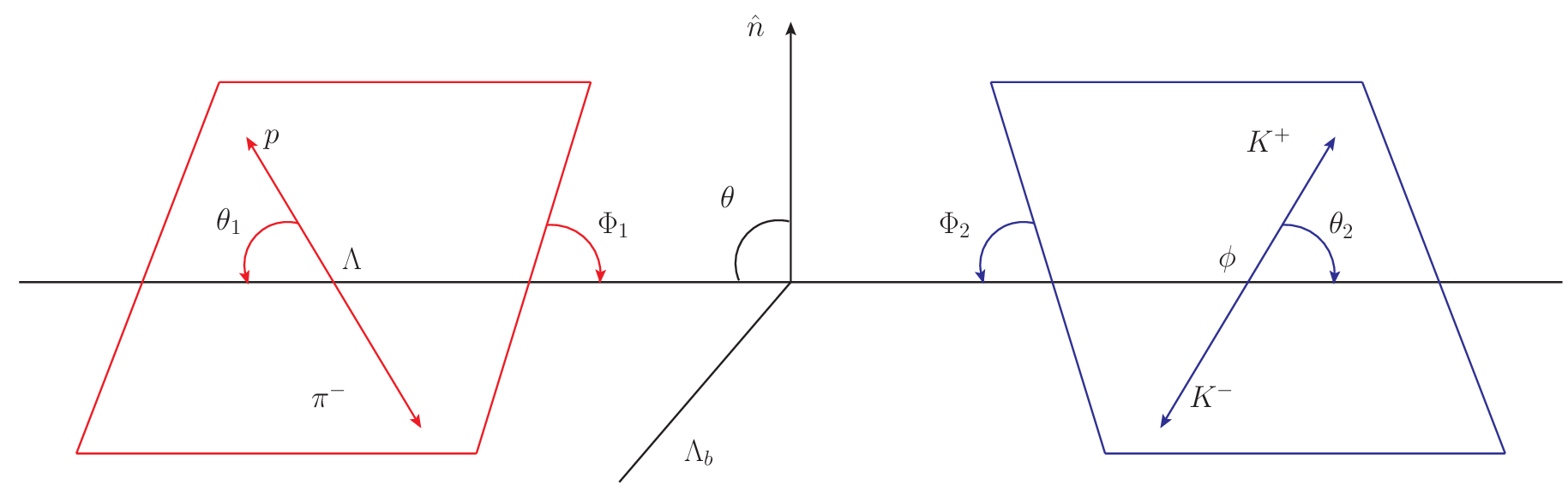}}
		\setlength{\abovecaptionskip}{1.0cm}
		\caption{Decay angles for the $\Lambda_b\rightarrow \Lambda (\rightarrow p\pi^-) \phi(\rightarrow K^+K^-)$ decay, where $\hat{n}$ is 
the normal unit-vector to the $\Lambda_b$ production plane. $\theta$ is the polar angle of the $\Lambda$ momentum with respect to $\hat{n}$ in the  $\Lambda_b$ rest-frame.
The angles $\theta_{1}$ and $\Phi_{1}$  are respectively the polar and azimuthal angles of the proton momentum in the $\Lambda$ rest frame
while $\theta_{2}$ and $\Phi_{2}$  are those of  $K^+$ in the $\phi$ rest frame.}
		\label{fig:angle}
	\end{center}
\end{figure}
Many asymmetry observables can be extracted from the angular distributions for the cascade decay $\Lambda_b\rightarrow \Lambda (\rightarrow p\pi^-) \phi(\rightarrow K^+K^-)$,
which can be parametrized  with three polar angles $\theta$, $\theta_1 $, and $\theta_2$ and two azimuthal angles $\Phi_1$ and $\Phi_2$  as illustrated in Fig.~\ref{fig:angle}.
The fivefold angular distribution   is written as~\cite{jhep111042021} 
\begin{eqnarray}\label{eq:omega}
\frac{1}{\Gamma}\frac{d^5\Gamma}{d\Omega}=N\sum_{\lambda,\lambda'=\pm \frac{1}{2}}(\frac{1}{2}+\lambda P_b)\left|\sum_{\lambda_\Lambda,\lambda_\phi}
A_{\lambda'}H_{\lambda_\Lambda,\lambda_\phi}
d^{\frac{1}{2}}_{\lambda',\lambda_\Lambda-\lambda_\phi}(\theta)d^{\frac{1}{2}}_{\lambda_\Lambda,\lambda}(\theta_1)d^1_{\lambda_\phi,0}(\theta_2)
e^{i(\lambda_\Lambda\phi_1+\lambda_\phi\phi_2)}\right|^2,
\end{eqnarray}
with $d\Omega=d\cos\theta d\cos\theta_1 d\cos\theta_2d\Phi_1d\Phi_2$.
The first angle, $\theta$, is the polar angle of the $\Lambda$ momentum in the $\Lambda_b$  rest-frame with respect to
the normal direction of the $\Lambda_b$ production plane.
The two sets of solid angles $(\theta_1, \Phi_1)$ and $(\theta_2, \Phi_2)$  describe the decays of the $\Lambda$ baryon  and the $\phi$ meson in their respective  rest-frame.
$d^j_{mm'} $  is the Wigner-d function, which can be expanded into a sequence of trigonometric functions.
$P_b$  describes the polarization of $\Lambda_b$.
$|A_{\lambda'}|^2=(\frac{1}{2}+\lambda' \alpha_\Lambda)$ with $\alpha_\Lambda$ the up-down asymmetry parameters for $\Lambda \rightarrow p\pi^-$~\cite{np15631}.
$N$ is an overall normalization factor
such that the integration of Eq.~(\ref{eq:omega}) over $d \Omega$ is equal to one.

Similar to the case of $B$ meson decays~\cite{Rui:2021kbn,Zhang:2021nlw}, a TPA $A_T$ represents an asymmetry between the decay rates $\Gamma$ with positive and negative values of $TP$,
\begin{eqnarray}
A_T\equiv \frac{\Gamma(TP>0)-\Gamma(TP<0)}{\Gamma(TP>0)+\Gamma(TP<0)},
\end{eqnarray}
where $TP$ denotes  a scalar triple product,
which is odd under the time reversal transformation ($T$),
and thus constitutes a potential signal of $CP$ violation assuming $CPT$ invariance.
Choosing the appropriate $TP$, one can construct the asymmetries as follows~\cite{Durieux:2016nqr}
\begin{eqnarray} \label{eq:ATs}
A_T^1&=& \frac{\Gamma(\cos \theta_2 \sin (\Phi_1+\Phi_2)>0 )-\Gamma(\cos \theta_2 \sin (\Phi_1+\Phi_2)<0 )}
{\Gamma(\cos \theta_2 \sin (\Phi_1+\Phi_2)>0 )+\Gamma(\cos \theta_2 \sin (\Phi_1+\Phi_2)<0 )}
=-\frac{\alpha_\Lambda}{\sqrt{2}}\frac{Im[H_{\frac{1}{2}0}H_{-\frac{1}{2}-1}^*+H_{-\frac{1}{2}0}H_{\frac{1}{2}1}^*]}{H_N},\nonumber\\
A_T^2&=& \frac{\Gamma(\cos \theta_2 \sin \Phi_2>0 )-\Gamma(\cos \theta_2 \sin \Phi_2<0 )}
{\Gamma(\cos \theta_2 \sin \Phi_2>0 )+\Gamma(\cos \theta_2 \sin \Phi_2<0 )}
=-\frac{P_b}{\sqrt{2}}\frac{Im[H_{-\frac{1}{2}-1}H_{-\frac{1}{2}0}^*+H_{\frac{1}{2}1}H_{\frac{1}{2}0}^*]}{H_N},\nonumber\\
A_T^3&=& \frac{\Gamma(\cos \theta \cos \theta_2\sin (\Phi_1+\Phi_2)>0 )-\Gamma(\cos \theta \cos \theta_2\sin (\Phi_1+\Phi_2)<0 )}
{\Gamma(\cos \theta \cos \theta_2\sin (\Phi_1+\Phi_2)>0 )+\Gamma(\cos \theta \cos \theta_2\sin (\Phi_1+\Phi_2)<0 )}
=\frac{P_b\alpha_\Lambda}{2\sqrt{2}}\frac{Im[H_{-\frac{1}{2}-1}H_{\frac{1}{2}0}^*-H_{\frac{1}{2}1}H_{-\frac{1}{2}0}^*]}{H_N},\nonumber\\
A_T^4&=& \frac{\Gamma(\cos \theta_1 \cos \theta_2\sin \Phi_2>0 )-\Gamma(\cos \theta_1 \cos \theta_2\sin \Phi_2<0 )}
{\Gamma(\cos \theta_1 \cos \theta_2\sin \Phi_2>0 )+\Gamma(\cos \theta_1 \cos \theta_2\sin \Phi_2<0 )}
=\frac{P_b\alpha_\Lambda}{2\sqrt{2}}\frac{Im[H_{-\frac{1}{2}-1}H_{-\frac{1}{2}0}^*-H_{\frac{1}{2}1}H_{\frac{1}{2}0}^*]}{H_N},\nonumber\\
A_T^5&=& \frac{\Gamma(\sin \Phi_1>0 )-\Gamma(\sin \Phi_1<0 )}{\Gamma(\sin \Phi_1>0 )+\Gamma(\sin \Phi_1<0 )}
=-\frac{P_b\pi\alpha_\Lambda}{4}\frac{Im[H_{-\frac{1}{2}0}H_{\frac{1}{2}0}^*]}{H_N},\nonumber\\
A_T^6&=& \frac{\Gamma(\sin (\Phi_1+2\Phi_2)>0 )-\Gamma(\sin (\Phi_1+2\Phi_2)<0 )}{\Gamma(\sin (\Phi_1+2\Phi_2)>0 )+\Gamma(\sin (\Phi_1+2\Phi_2)<0 )}
=-\frac{P_b\pi\alpha_\Lambda}{4}\frac{Im[H_{\frac{1}{2}1}H_{-\frac{1}{2}-1}^*]}{H_N}.
\end{eqnarray}
Then the true asymmetries are defined as~\cite{plb538309,Gronau:2015gha}
\begin{eqnarray}
A^i_T(true)=\frac{A_T^i-\bar{A}_T^i}{2},
\end{eqnarray}
where $\bar{A}_T^i$ with $i=1-6$ are the corresponding quantities for the  charge-conjugate process.
A significant deviation from zero in these observables would indicate $CP$ violation.
Note that the above six asymmetries have the same forms as those in Ref.~\cite{jhep111042021},
apart from the additional multiplicative factors derived from the integral of the decay angles.
Changing the sine of the azimuthal angle in above $TP$ to cosine, we can obtain six new asymmetries,
which are proportional to the real part of bilinear combinations of the helicity amplitudes.
The corresponding expressions can be obtained from Eq.~(\ref{eq:ATs}) with the replacement of $Im\rightarrow Re$.
Similarly, combining the quantities for the corresponding $CP$-conjugate process,
they have the form of $Re[H'H^*-\bar{H}'\bar{H}^*]$,
 which is characteristic of direct $CP$ asymmetry.

Integrating Eq.~(\ref{eq:omega}) over the two azimuthal angles $\Phi_1$ and $\Phi_2$,
the angular distribution is reduced  to eight terms as shown in~\cite{plb72427} (see Table 2),
which can be written in terms of the following three independent angular observables
\begin{eqnarray}\label{eq:alp}
\alpha_b &=&-|\hat{H}_{\frac{1}{2}1}|^2+|\hat{H}_{-\frac{1}{2}-1}|^2+|\hat{H}_{\frac{1}{2}0}|^2-|\hat{H}_{-\frac{1}{2}0}|^2, \nonumber\\
r_0 &=&|\hat{H}_{\frac{1}{2}0}|^2+|\hat{H}_{-\frac{1}{2}0}|^2, \nonumber\\
r_1 &=&|\hat{H}_{\frac{1}{2}0}|^2-|\hat{H}_{-\frac{1}{2}0}|^2,
\end{eqnarray}
with $|\hat{H}_{\lambda_\Lambda\lambda_\phi}|^2=|H_{\lambda_\Lambda\lambda_\phi}|^2/H_N$.
The asymmetry parameter $\alpha_b$ characterizes parity nonconservation in a weak decay of $\Lambda_b$.
 $r_0 $ and $r_1$ are the longitudinal unpolarized and polarized parameters, respectively.

 Further integrating $(\theta,\theta_2)$ and $(\theta,\theta_1)$, respectively, one can extract two more interesting asymmetries
 \begin{eqnarray}\label{eq:alp12}
\alpha_{\lambda_\Lambda}&=&|\hat{H}_{\frac{1}{2}0}|^2+|\hat{H}_{\frac{1}{2}1}|^2-|\hat{H}_{-\frac{1}{2}-1}|^2-|\hat{H}_{-\frac{1}{2}0}|^2, \nonumber\\
\alpha_{\lambda_\phi}&=&|\hat{H}_{\frac{1}{2}0}|^2+|\hat{H}_{-\frac{1}{2}0}|^2-|\hat{H}_{\frac{1}{2}1}|^2-|\hat{H}_{-\frac{1}{2}-1}|^2,
\end{eqnarray}
 where the former describes the polarization of $\Lambda$ baryon,
 and the latter represents the asymmetry between the longitudinal and transverse polarizations of $\phi$ meson.
We see that $\alpha_{\lambda_\Lambda}$ and $\alpha_{\lambda_\phi}$ are odd and even under parity transformation, respectively.

\section{Numerical results}\label{sec:results}
Numerical results on the invariant and helicity amplitudes, branching ratio, angular observables, direct $CP$ violation, and TPAs will be presented in this section.
In our numerical study, the values of relevant masses (GeV), lifetimes (ps), and the Wolfenstein parameters for the CKM matrix
are taken from the latest experimental values~\cite{pdg2020}:
\begin{eqnarray}
M&=&5.6196, \quad  m_{\Lambda}=1.116,  \quad m_b=4.8, \quad  m_\phi=1.019, \nonumber\\
\quad \tau&=&1.464, \quad \lambda =0.22650, \quad  A=0.790,  \quad \bar{\rho}=0.141, \quad \bar{\eta}=0.357.
\end{eqnarray}
Other nonperturbative parameters appearing in the hadron LCDAs have been specified in the preceding section.

\begin{table}
\footnotesize
\caption{ Contributions from the various topologies to the invariant amplitudes $(10^{-10})$ for $\Lambda_b\rightarrow \Lambda \phi$ decay.
The labels $P$, $PE(PB)$, and $E(B)$ corresponds to the contributions from the penguin emission, penguin exchange, and the  $W$ exchange diagrams, respectively,
and the last column is their sum. Only central values are presented here.}
\label{tab:amp}
\begin{tabular}[t]{lcccccc}
\hline\hline
Amplitude & $P$           & $PE$           & $E$           & $PB$          & $B$           & Total          \\ \hline
$A^L_1$   & $-0.59-i3.90$ & $4.24-i10.82$  & $1.14+i0.71$  & $0.39+i0.02$  & $-0.02+i0.05$ & $5.16-i13.94$  \\
$B^L_1$   & $-1.69+i4.05$ & $-3.86+i12.20$ & $-1.46-i0.83$ & $-0.56+i0.00$ & $0.01-i0.07$  & $-7.55+i15.36$ \\
$A^L_2$   & $0.43-i36.09$ & $9.45-i25.39$  & $2.66+i1.65$  & $1.47+i0.02$  & $-0.04+i0.14$ & $13.97-i59.67$ \\
$B^L_2$   & $2.75-i15.10$ & $7.92-i19.70$  & $2.59+i1.28$  & $0.64+i0.01$  & $-0.03+i0.13$ & $13.87-i33.39$ \\
$A_1^T$   & $-0.53-i2.54$ & $2.22-i4.70$   & $0.64+i0.59$  & $0.14-i0.11$  & $0.00+i0.03$  & $2.47-i6.73$   \\
$B_1^T$   & $-1.99+i2.73$ & $-2.00+i5.45$  & $-1.08-i0.69$ & $-0.20+i0.21$ & $0.00-i0.04$  & $-5.26+i7.65$  \\
\hline\hline
\end{tabular}
\end{table}

As stressed before, all the five topologies  can be evaluated systematically in PQCD.
It is interesting to compare the relative strengths among these topologies,
whose results are shown separately in Table~\ref{tab:amp}.
The labels $P$, $PE(PB)$, and $E(B)$ corresponds to the contributions from the penguin emission, penguin exchange, and $W$ exchange,  respectively,
and the last column is their sum.
It is found that the decay amplitudes are governed by the penguin emission $P$ and penguin exchange diagrams $PE$,
which contribute at the  same order of magnitude,
while the $W$ exchange ones suffer from severe CKM suppression and are smaller by one or two orders of magnitude.
The contributions from $B$ and $PB$-type exchange diagrams are predicted to be vanishingly small.

\begin{table}
\footnotesize
\caption{The contributions from $\Phi^V $, $\Phi^A $, and $\Phi^T $ in the $\Lambda$ baryon LCDAs to the invariant amplitudes.}
\label{tab:ampvat}
\begin{tabular}[t]{lccc}
\hline\hline
Amplitude & $\Phi^V $                                  & $\Phi^A $                                     & $\Phi^T $                                    \\ \hline
$A^L_1$   & $1.3\times 10^{-13}	+i7.5\times 10^{-11}$  & $5.7\times 10^{-10}	-i1.2\times 10^{-9}$   & $-5.0\times 10^{-11}-i2.8\times 10^{-10}$    \\
$B^L_1$   & $4.5\times 10^{-11}	-i1.3\times 10^{-10}$  & $-9.2\times 10^{-10}+i1.3\times 10^{-9}$      & $1.2\times 10^{-10}	+i3.7\times 10^{-10}$ \\
$A^L_2$   & $-2.5\times 10^{-10}+i6.4\times 10^{-11}$  & $1.3\times 10^{-9}	-i4.3\times 10^{-9}$       & $4.0\times 10^{-10}	-i1.8\times 10^{-9}$  \\
$B^L_2$   & $9.0\times 10^{-11}	+i9.4\times 10^{-11}$  & $1.5\times 10^{-9}	-i3.5\times 10^{-9}$       & $-2.1\times 10^{-10}+i3.4\times 10^{-11}$    \\
$A_1^T$   & $-2.8\times 10^{-11}+i1.0\times 10^{-11}$  & $2.9\times 10^{-10}	-i5.0\times 10^{-10}$  & $-1.3\times 10^{-11}-i1.8\times 10^{-10}$    \\
$B_1^T$   & $1.2\times 10^{-10}	+i2.9\times 10^{-13}$  & $-6.7\times 10^{-10}+i5.4\times 10^{-10}$     & $2.1\times 10^{-11}	+i2.3\times 10^{-10}$ \\
\hline\hline
\end{tabular}
\end{table}

 In order to understand the mechanism responsible for the large contribution from the $PE$ diagrams,
 one has to   compare the relative contributions from various components in the LCDAs.
 In this study we take into account  the contributions from the $\Lambda_b$ baryon LCDAs up to twist 4 and the $\phi$ meson LCDAs up to twist 3,
 while the $\Lambda$ one is restricted to the leading-twist accuracy.
 We first compare in Table~\ref{tab:ampvat} the values of the invariant amplitudes from three components of the leading-twist LCDAs of $\Lambda$ baryon.
 It is observed that
the mainly contributions come from the component $\Phi^A $   as a consequence of the symmetry relations in Eq.~(\ref{eq:vat}).
Conversely, contributions from the components $\Phi^V$ and $\Phi^T$ are relative small due to the antisymmetry under the interchange of $x$ and $1-x$.
In particular, the contribution of $\Phi^V$ is further suppressed by the small coefficients of the terms $x_2^2-x_3^2$ and $x_2-x_3$ in Eq.~(\ref{eq:vat}).
Similar feature  has also been observed in previous PQCD calculations on the $\Lambda_b\rightarrow \Lambda J/\psi$ mode~\cite{prd65074030,220604501}.
Contributions from twist-2 and  twist-3 LCDAs of $\phi$ meson are also displayed separately in Table~\ref{tab:ampphi}.
It can be shown that their contributions are comparable.
The important twist-3 meson LCDAs contributions were also observed in the penguin-dominated $B$ meson decay, such as  $B_s \rightarrow \phi\phi$~\cite{prd76074018}.

\begin{table}
\footnotesize
\caption{The values of invariant amplitude from twist-2 and twist-3  LCDAs of $\phi$ meson  for $\Lambda_b\rightarrow \Lambda \phi$ decays.}
\label{tab:ampphi}
\begin{tabular}[t]{lcc}
\hline\hline
Amplitude & twist-2                                   & twist-3                                   \\ \hline
$A^L_1$   & $1.0\times 10^{-10}	-i2.0\times 10^{-10}$ & $4.1\times 10^{-10}	-i1.2\times 10^{-9}$  \\
$B^L_1$   & $-3.9\times 10^{-10}+i2.3\times 10^{-10}$ & $-3.7\times 10^{-10}+i1.3\times 10^{-9}$  \\
$A^L_2$   & $3.9\times 10^{-10}	-i3.5\times 10^{-9}$  & $1.0\times 10^{-9}	-i2.4\times 10^{-9}$  \\
$B^L_2$   & $4.7\times 10^{-10}	-i9.9\times 10^{-10}$ & $9.2\times 10^{-10}	-i2.3\times 10^{-9}$  \\
$A_1^T$   & $2.9\times 10^{-10}	-i5.0\times 10^{-10}$ & $-4.7\times 10^{-11}-i1.8\times 10^{-10}$ \\
$B_1^T$   & $-2.5\times 10^{-10}+i5.7\times 10^{-10}$ & $-2.8\times 10^{-10}+i2.0\times 10^{-10}$ \\
\hline\hline
\end{tabular}
\end{table}

Given this situation,
it is expected that the dominant contribution to the invariant amplitudes comes from the combination of
$\Phi^A $ and twist-2 LCDAs of $\Lambda_b$, such as the $\Phi^A \Psi_2 $ term.
Note that the invariant amplitudes  involve different sets of $\Lambda_b$ baryon LCDAs through different Feynman diagrams.
 In the following analysis we take diagrams of $P_{c7,d6}$ and $PE_{b3}$ as examples to illustrate,
 which dominate the $P$ and $PE$-type amplitudes, respectively.
As can be seen from Table~\ref{tab:amppc} that the dominant term $\Phi^A \Psi_2$ contributes to $A_1,B_1$ and $A_2,B_2$ through the nonfactorizable diagrams $P_{d6}$ and $P_{c7}$, respectively.
Thus the penguin emission decay amplitudes are governed by the twist-2 contributions.
Nevertheless, for the penguin exchange diagram $PE_{b3}$, the twist-2 term vanish at the level of the theoretical accuracy in the current formalism.
The leading contribution to $PE_{b3}$ comes from the twist-4 term, $\Phi^A \Psi_4$.
As stated in~\cite{220204804}, the higher-twist effects are crucial at the realistic scale of the $b$ quark mass,
and the contribution from twist-4 $\Lambda_b$ baryon LCDAs could overcome the power suppression from  $1/M$
with respect to  the twist-2 one due to the enhancement from the endpoint region.
It is not surprising that the  twist-2 and twist-4 contributions are of the same order for a similar reason in this work.
Therefore we can explain why  the penguin exchange amplitudes  are  in fact 
 at the same order as the penguin emission ones, which can be seen in the columns of $P$ and $PE$ of Table~\ref{tab:amp} .

Table~\ref{tab:amptw} indicates clearly that the higher-twist contributions significantly enhance the penguin exchange amplitudes.
 The picture of  significant exchange  topological contributions differs from the case of the color-allowed decays of $\Lambda_b \rightarrow \Lambda_c \pi, \Lambda_c K$~\cite{prd105073005},
in which the emission type amplitude  is dominant and accounts for more than $90\%$ of the total decay amplitudes.
 We should point out, however, that  only the leading-twist LCDAs of the $\Lambda_b$ baryon was considered in~\cite{prd105073005}.
The hierarchy relationship may be modified to some extent if the higher-twist contributions are included, which is an intriguing topic for future research.

\begin{table}
\footnotesize
\caption{The values of the invariant amplitudes from various twists of the $\Lambda_b$ baryon LCDAs for $P$ and $PE$ diagrams.}
\label{tab:amptw}
\begin{tabular}[t]{lccc}
\hline\hline
Amplitude & Twist & $P$                                         & $PE$                                          \\ \hline
$     $   & 2     & $6.0\times 10^{-11}	-i1.3\times 10^{-10}$   & $2.8\times 10^{-11}	+i6.5\times 10^{-11}$   \\
$A^L_1$   & 3     & $-1.0\times 10^{-10}-i4.9\times 10^{-11}$   & $4.4\times 10^{-11}	-i1.6\times 10^{-10}$   \\
$     $   & 4     & $-1.7\times 10^{-11}-i2.1\times 10^{-10}$   & $3.5\times 10^{-10}	-i9.9\times 10^{-10}$   \\ \hline
$     $   & 2     & $-6.7\times 10^{-11}+i1.3\times 10^{-10}$   & $-2.8\times 10^{-11}	-i9.0\times 10^{-11}$   \\
$B^L_1$   & 3     & $1.6\times 10^{-10}	-i3.0\times 10^{-11}$   & $5.3\times 10^{-11}	+i4.2\times 10^{-10}$   \\
$     $   & 4     & $-2.6\times 10^{-10}+i3.0\times 10^{-10}$   & $-4.1\times 10^{-10}	+i8.9\times 10^{-10}$   \\ \hline
$     $   & 2     & $2.5\times 10^{-10}	-i2.2\times 10^{-9}$    & $9.8\times 10^{-11}	+i2.2\times 10^{-10}$   \\
$A^L_2$   & 3     & $-7.1\times 10^{-11}-i1.1\times 10^{-9}$    & $7.3\times 10^{-11}	-i8.5\times 10^{-10}$   \\
$     $   & 4     & $-1.3\times 10^{-10}-i2.6\times 10^{-10}$   & $7.7\times 10^{-10}	-i1.9\times 10^{-9}$    \\ \hline
$     $   & 2     & $1.8\times 10^{-10}	-i1.5\times 10^{-9}$    & $2.7\times 10^{-11}	+i1.1\times 10^{-10}$   \\
$B^L_2$   & 3     & $-2.7\times 10^{-10}+i4.0\times 10^{-10}$   & $1.1\times 10^{-10}	-i3.9\times 10^{-10}$   \\
$     $   & 4     & $3.6\times 10^{-10}	-i4.4\times 10^{-10}$   & $6.5\times 10^{-10}	-i1.7\times 10^{-9}$    \\ \hline
$     $   & 2     & $2.5\times 10^{-11}	-i5.9\times 10^{-11}$   & $-1.7\times 10^{-11}	+i2.9\times 10^{-11}$   \\
$A_1^T$   & 3     & $-1.0\times 10^{-10}-i7.7\times 10^{-12}$   & $4.3\times 10^{-11}	-i1.6\times 10^{-10}$   \\
$     $   & 4     & $2.3\times 10^{-11}	-i1.9\times 10^{-10}$   & $2.0\times 10^{-10}	-i3.4\times 10^{-10}$   \\ \hline
$     $   & 2     & $-3.3\times 10^{-11}+i6.2\times 10^{-11}$   & $2.5\times 10^{-11}	-i4.6\times 10^{-11}$   \\
$B_1^T$   & 3     & $1.8\times 10^{-10}	+i2.0\times 10^{-12}$   & $2.3\times 10^{-11}	+i2.9\times 10^{-10}$   \\
$     $   & 4     & $-3.4\times 10^{-10}+i2.1\times 10^{-10}$   & $-2.5\times 10^{-10}	+i3.0\times 10^{-10}$   \\
\hline\hline
\end{tabular}
\end{table}

The factorizable amplitudes contribute only via the penguin emission diagrams,
such as $P_{a1-a5}$, $P_{b1-b5}$, and $P_{e1,e2,f1,f2}$ as shown in Fig.~\ref{fig:FeynmanT},
in which the $s\bar s$ of $\phi$ mesons are created by weak vertices.
In Table~\ref{tab:facamp}, we present the factorizable and
nonfactorizable contributions in the decay amplitudes.
It is observed that the nonfactorizable contributions dominate over the factorizable ones,
which is similar to the cases of the decay of $\Lambda_b\rightarrow \Lambda J/\psi$~\cite{prd65074030,220604501} and
 $\Lambda_b\rightarrow p\pi,pK$~\cite{prd80034011}.

\begin{table}
\footnotesize
\caption{The values of the invariant amplitudes from the factorizable and nonfactorizable diagrams for $\Lambda_b\rightarrow \Lambda \phi$   decay.
}
\label{tab:facamp}
\begin{tabular}[t]{lcccc}
\hline\hline
Amplitude & Factorizable           & Nonfactorizable                             \\ \hline
$A^L_1$   & $-1.5\times 10^{-11} $ & $5.3\times 10^{-10} -i1.4\times 10^{-9}   $ \\
$B^L_1$   & $-2.7\times 10^{-11} $ & $-7.3\times 10^{-10} +i1.5\times 10^{-9}  $ \\
$A^L_2$   & $-6.1\times 10^{-11} $ & $1.5\times 10^{-9} -i6.0\times 10^{-9}    $ \\
$B^L_2$   & $6.3\times 10^{-11}  $ & $1.3\times 10^{-9} -i3.3\times 10^{-9}    $ \\
$A_1^T$   & $-1.5\times 10^{-11} $ & $2.6\times 10^{-10} -i6.7\times 10^{-10}  $ \\
$B_1^T$   & $-3.1\times 10^{-11} $ & $-5.0\times 10^{-10} +i7.7\times 10^{-10} $ \\
\hline\hline
\end{tabular}
\end{table}

\begin{table}[!htbh]
	\caption{Helicity amplitudes and  phases of $\Lambda_b\rightarrow \Lambda \phi$ decay. The last row corresponds to the magnitude squared of normalized helicity amplitudes.}
	\label{tab:helicity}
	\begin{tabular}[t]{lcccc}
		\hline\hline
		$\lambda_{\Lambda}\lambda_{\phi}$  &$\frac{1}{2}1$ & $-\frac{1}{2}-1$ & $\frac{1}{2}0$ & $-\frac{1}{2}0$ \\ \hline		
		$H_{\lambda_{\Lambda}\lambda_{\phi}}$ & $6.6\times10^{-10}+i1.1\times10^{-9}$  & $4.0\times10^{-9}-i7.9\times10^{-9}$  & $8.6\times10^{-10}+i9.9\times10^{-9}$   & $1.6\times10^{-9}-i2.8\times10^{-8}$   \\	
		$\Phi_{\lambda_{\Lambda}\lambda_{\phi}}$ & $1.04$  & $-1.10$  & $1.48$   & $-1.51$  \\		
		$|\hat{H}_{\lambda_{\Lambda}\lambda_{\phi}}|^2$  &$2.0\times10^{-3}$  &$8.3\times10^{-2}$  &$0.105$  &$0.81$\\
		\hline\hline
	\end{tabular}
\end{table}	

By using the results of  Table~\ref{tab:amp}, one can calculate the helicity amplitudes according to Eq.~(\ref{eq:helicity}),
whose numerical results are displayed in the Table~\ref{tab:helicity}.
It is observed that the amplitudes are dominated by $H_{-\frac{1}{2}0}$ which occupies about $81\%$ of the full contribution.
It implies the negative-helicity component of the $\Lambda$ baryon and longitudinally polarized $\phi$ meson in the final states are preferred. 
Contributions from the transverse polarizations of $\phi$ meson are rather small, which amounts to less than $10\%$.
As pointed out in~\cite{prd66094004}, the transverse amplitudes are suppressed relative to the longitudinal ones  by a factor $r_{\phi}$,
which can also be seen from Eq.~(\ref{eq:helicity}).
This situation differs from the case of $\Lambda_b\rightarrow \Lambda J/\psi$ decays~\cite{220604501},
where  their contributions are comparable because the $J/\psi$ is very heavy (three times as lager as the $\phi$),
and thus the suppression is not obvious.
The domination of $H_{-\frac{1}{2}0}$ is consistent with the expectation from the heavy-quark limit and the left-handed nature of the weak interaction
and  the prediction in GFA~\cite{jhep111042021}.
As shown in~\cite{jhep111042021},
the four helicity amplitudes in the GFA share the same complex phase, which come from the common effective Wilson coefficients.
However, in our calculation, the nonfactorizable contributions  are the major source of the strong phases,
 and each helicity amplitude has a different phase as exhibited in Table~\ref{tab:facamp}.

\begin{table}[!htbh]
	\caption{Branching ratios  and  asymmetry parameters for the $\Lambda_b\rightarrow \Lambda \phi$ decay.
The theoretical errors correspond to the uncertainties due to $\omega=0.40\pm0.04$ GeV and the hard scale $t=(1.0\pm0.2)t$, respectively.}
	\label{tab:branching}
	\begin{tabular}[t]{lccccc}
		\hline\hline
    $\mathcal{B}$      & $\alpha_b$ & $\alpha_{\lambda_\Lambda}$ & $\alpha_{\lambda_\phi}$  &$r_0$  &$r_1$ \\ \hline		
    $6.9^{+1.9+1.8}_{-2.0-1.6}\times10^{-6}$ & $-0.63^{+0.16+0.00}_{-0.01-0.03}$    & $-0.79^{+0.12+0.00}_{-0.02-0.04}$
       & $0.83^{+0.03+0.01}_{-0.03-0.05}$      &$0.91^{+0.01+0.01}_{-0.02-0.02}$ & $-0.71^{+0.14+0.00}_{-0.01-0.02}$ \\
		\hline\hline
	\end{tabular}
\end{table}

After obtaining the values of the helicity amplitudes,
we can compute the branching ratio and various asymmetries through Eqs.~(\ref{eq:two}), (\ref{eq:alp}), and (\ref{eq:alp12}).
The numerical results are collected in Table~\ref{tab:branching},
where the first and second uncertainties arise
from the shape parameter $\omega_0=0.40\pm0.04$ GeV in the $\Lambda_b$ baryon LCDAs
and hard scale $t$ varying from  $0.8t$  to $1.2t$, respectively. We draw the following observations:
\begin{enumerate}
\item
The PQCD prediction of the branching ratio involves a large uncertainty because of the sensitivity to the nonperturbative hadronic LCDAs,
which are of limited accuracy due to our lack of understanding of QCD dynamics at low energies.
Pinning down the uncertainties of LCDAs is an essential prescription to improve the accuracy of PQCD calculations.
In contrast to the  branching ratio,
most of the asymmetry observables  are insensitive to the nonperturbative QCD effects
because the resulting uncertainties have been eliminated in the ratios,
indicates they can serve as the ideal quantities to test the PQCD approach.
\item
The predicted branching ratio is slightly larger than the LHCb  measured value $(5.18\pm 1.29)\times 10^{-6}$~\cite{plb759282},
where multiple uncertainties are added in quadrature.
However, recent updates from the PDG on the web and the world average from the Heavy Flavor Averaging Group (HFLAV) give
$(9.8\pm2.6)\times10^{-6}$ and $(10.1^{+2.9}_{-2.5})\times10^{-6}$~\cite{HFLAV:2022pwe}, respectively.
Both of the central values
are larger than the LHCb data by a factor of two.
We notice that in fact all these data come from the measurement of the relative branching fraction of
$\Lambda_b\rightarrow \Lambda \phi$ to $B^0\rightarrow K^0 \phi$ by LHCb~\cite{plb759282},
and the discrepancies are mainly caused by using different values for the production rate ratio $f_{\Lambda_b}/f_d$
with $f_{\Lambda_b}(f_d)$ being the fragmentation fractions of $b$ quarks to $\Lambda_b(B^0)$.
\item
By comparison, the estimates based on GFA
gives $\mathcal{B}(\Lambda_b\rightarrow \Lambda \phi)=(1.77^{+1.76}_{-1.71}\pm 0.24)\times10^{-6}$~\cite{epjc76399},
which suffers large  uncertainties from the nonfactorizable effects.
With the number of colors setting as $N^{eff}=2$, the value can be enhanced to   $(3.53\pm 0.24)\times10^{-6}$,
which is  still half of our prediction and apparently lower than the world averages.
The result from the QCDF~\cite{prd99054020}, yields $(6.33^{+0.60+1.57+0.83}_{-0.68-1.56-0.61})\times10^{-7}$,
 is below our number by roughly one order of magnitude.
The smallness is ascribed to the important nonfactorizable effects, such as hard spectator interactions and power corrections,
were not included in their calculations.
\item
The up-down asymmetry  $\alpha_b$ is predicted to be $-0.63^{+0.16+0.00}_{-0.01-0.03}$  in PQCD, away from -1
because the considered process also receives sizeable contributions from the QCD penguin operators $O_{5,6,7,8}$ via the $V+A$ current.
Our central value is somewhat larger than the QCDF calculation $-0.8$ presented in~\cite{prd99054020}.
The asymmetry parameters of $\alpha_{\lambda_\Lambda}$ and $\alpha_{\lambda_\phi}$ evaluated in GFA are $-0.99$ and $0.86$~\cite{jhep111042021}, respectively,
which are comparable to our results in Table~\ref{tab:branching}.
The obtained  predictions on the $r_0$ and $r_1$, received less theoretical and experimental attentions, can be compared in future. 

\end{enumerate}

We now discuss the TPAs in $\Lambda_b\rightarrow \Lambda \phi$ decay.
From Eq.~(\ref{eq:ATs}), one can see that two parameters $\alpha_\Lambda$ and $P_b$ enter into the angular distribution and appear in the TPAs.
For the former, we use  the new experimental PDG average values $\alpha_\Lambda=0.732\pm0.014$ and $\alpha_{\bar{\Lambda}}=-0.758\pm0.012$~\cite{pdg2020},
 deduced from the measurements by the BES-III~\cite{np15631} and CLAS~\cite{Ireland:2019uja} Collaborations.
The value of $P_b$  is annoying because it depends on the production mechanism of the $\Lambda_b$.
Here we assume that $\Lambda_b$ baryons are produced directly  in the hadron collisions,
where the longitudinal polarization is expected to vanish due to parity conservation in strong interactions~\cite{LHCb:2020iux}.
The $CP$ invariance between the $\Lambda_b$  and $\bar{\Lambda}_b$ angular distributions implies
the relation of $\bar{P}_b=-P_b$ can be used~\cite{CMS:2018wjk}.
Experimentally,
the polarization had been measured from the angular distributions of $\Lambda_b \rightarrow \Lambda J/\psi$ decay
by the LHCb~\cite{LHCb:2020iux}, CMS~\cite{CMS:2018wjk}  and ATLAS~\cite{ATLAS:2014swk}  experiments,
which have yielded a value consistent with zero,
though polarization of $10\%$ is possible given statistical uncertainties~\cite{plb72427}.
Theoretically, it has been suggested that
the value of $P_b$ would reach up to the $(10-20)\%$ level~\cite{plb614165,Hiller:2007ur}.
In the following analysis, we take  $P_b=0.1$  as a rough numerical  estimates.

\begin{table}
\footnotesize
\caption{PQCD predictions for the TPAs with $P_b=0.1$.
The sources of the theoretical errors are the same as in Table~\ref{tab:branching} but added in quadrature.}
\label{tab:tpas}
\begin{tabular}[t]{lccc}
\hline\hline
        & $A_T^i$   & $\bar{A}_T^i$  & $A_T^i(true)$      \\ \hline
i=1   &$-1.4^{+4.5}_{-0.1}\times 10^{-2}$ &$1.3^{+2.0}_{-3.7}\times 10^{-2}$	      &$-1.4^{+4.1}_{-1.1}\times 10^{-2}$  \\
i=2   &$-6.9^{+0.0}_{-0.4}\times 10^{-3}$ &$-3.5^{+2.8}_{-3.6}\times 10^{-3}$	      &$-1.7^{+1.8}_{-1.6}\times 10^{-3}$  \\
i=3   &$-1.8^{+2.2}_{-0.6}\times 10^{-3}$ &$-5.7^{+0.0}_{-1.7}\times 10^{-4}$	      &$-0.6^{+1.2}_{-0.3}\times 10^{-3}$  \\
i=4   &$2.8^{+2.3}_{-0.0}\times 10^{-3}$ &$1.8^{+2.0}_{-1.6}\times 10^{-3}$	      &$0.5^{+2.0}_{-1.0}\times 10^{-3}$  \\
i=5   &$2.4^{+0.0}_{-1.8}\times 10^{-3}$ &$-3.6^{+1.2}_{-0.2}\times 10^{-3}$	      &$3.0^{+0.1}_{-1.5}\times 10^{-3}$  \\
i=6   &$-5.9^{+0.0}_{-0.4}\times 10^{-4}$ &$-5.5^{+0.0}_{-0.1}\times 10^{-4}$	      &$-0.2^{+0.1}_{-0.2}\times 10^{-4}$  \\
\hline\hline
\end{tabular}
\end{table}

The calculated TPAs in GFA must be zero~\cite{jhep111042021} since their helicity amplitudes  share the same complex phase as mentioned above.
In contrast, in PQCD regime,  the helicity amplitudes always have different strong phases, which mainly originate from the nonfactorizable diagrams as shown in Table~\ref{tab:facamp}.
This will lead to nonzero TPAs, which  are summarized in Table~\ref{tab:tpas}.
As expected the PQCD results on these asymmetries come out to be quite small, range from $10^{-4}$ to $10^{-2}$.
$A_T^1$  do not involve the polarization of the initial state and could be measured in the unpolarized angular distribution.
Its relative large value of order $10^{-2}$ makes it is the best candidate to look for in experimental searches.
The small values of $A_T^3$ and $A_T^4$  are mainly attributed by a substantial cancellation of the two component as can be seen in Eq.~(\ref{eq:ATs}).
The difference between $A_T^2$ and $A_T^4$ is ascribed to the distinct combinations of
$H_{-\frac{1}{2}-1}H_{-\frac{1}{2}0}^*$ and $H_{\frac{1}{2}1}H_{\frac{1}{2}0}^*$ in Eq.~(\ref{eq:ATs}).
The smallness of $A_T^6$ can be traced to the transverse polarization components suffer the power suppression as explained above.
In addition,  the $W$ exchange amplitudes associated with the CKM matrix elements $V_{ub}V^*_{us}$ provide the weak phase,
which interfere with the penguin amplitudes to produce nonvanishing true TPAs  as given in the last column of Table~\ref{tab:tpas}.
As noted previously, $W$ exchange contributions are highly CKM suppressed relative to the penguin ones,
most the true TPAs are estimated to be less than $\mathcal{O}(1\%)$ in magnitude,
compatible with the absence of $CP$ violation.
We therefore conclude that any measurement of a sizeable TPAs in the decay
is an unequivocal signal of new physics.

The partial angular analysis of  the considered process  has been studied by LHCb~\cite{plb759282},
in which four asymmetries are measured to be consistent with zero.
Note that these asymmetries are defined by the  so-called special angles~\cite{Leitner:2006sc},
which are explicitly given in Ref.~\cite{Durieux:2016nqr}.
As stated in~\cite{Durieux:2016nqr}, such angular asymmetries give access to terms proportional to the off-diagonal elements of the $\Lambda_b$ polarization density matrix,
which are absent in the current case since the $\Lambda_b$ is assumed to be produced by the strong interaction which preserves parity.
Therefore all the four asymmetries vanish in our calculations.

Finally, we predict the direct $CP$ asymmetry of the  $\Lambda_b \rightarrow \Lambda \phi$ decay, which is defined by
\begin{eqnarray}
A_{CP}=\frac{\Gamma(\Lambda_b \rightarrow \Lambda \phi)-\Gamma(\bar{\Lambda}_b \rightarrow \bar{\Lambda} \phi)}
{\Gamma(\Lambda_b \rightarrow \Lambda \phi)+\Gamma(\bar{\Lambda}_b \rightarrow \bar{\Lambda} \phi)},
\end{eqnarray}
where the overline denotes the antiparticles.
It is known that at least two amplitudes with nontrivial relative strong and weak phases are required to produce a nonvanishing direct $CP$ violation.
As already remarked above, the tree amplitudes contribute via the $W$ exchange diagrams,
while the penguin ones exist in both penguin emission and exchange diagrams.
The direct $CP$  asymmetry arises from the interference between the tree and penguin amplitudes.
Since  the tree-level $W$ exchange contribution is suppressed by the CKM matrix elements $|V_{ub}V_{us}^*/V_{tb}V_{ts}^*|\sim 0.02$ compared with the penguin ones,
the resulting  direct $CP$ asymmetry would be very small with the value of $-1.0^{+1.0}_{-1.5}\%$,
where the sources of the theoretical errors are the same as in Table~\ref{tab:branching} but added in quadrature.
The corresponding value from QCDF and GFA are $1.6^{+0.4}_{-0.3}\%$~\cite{prd99054020} and $1.4^{+0.7}_{-0.1}\%$~\cite{prd95093001}, respectively.
Our central value agrees with theirs in magnitude but differs in sign.
These results can be checked by future experiments.

\section{conclusion}\label{sec:sum}
The weak decay of the bottom baryon provides useful information about the strong interaction and serves as an important probe for testing various theoretical approaches.
Because baryons are three-quark systems in the conventional quark model,
their weak decays contain  a large number of topological diagrams,
making QCD dynamics involved in the hadronic matrix element extremely complicated.
Nonfactorizable contributions, in particular the exchange topological contributions,
are more difficult to evaluate from first principles.
The perturbative QCD approach is a powerful tool to analyze the heavy baryon decays,
in which both the emission and exchange topologies can be evaluated systematically.
In this work, we have carried out a systematic study on the penguin-dominant $\Lambda_b\rightarrow \Lambda \phi$ decay,
whose branching ratio was predicted to be much small  in previous literature
with  only the emission topological contributions being taken into account.
Since the PQCD calculations of the baryon decay start at two order of $\alpha_s$,
 $\Lambda_b\rightarrow \Lambda \phi$ proceeds simultaneously through the charged current $b \rightarrow u$ transition and neutral-current $b\rightarrow s $ transition.
The former corresponds to the tree diagrams contribution resulting from the $W$ exchange diagrams,
while the latter belongs to the penguin contributions,
which could be further cataloged into penguin emission and penguin exchange ones.
All the possible Feynman diagrams can be classified into five topological types, namely  $P$, $PE$, $E$, $B$, and $PB$, respectively.
It is observed that the decay amplitudes are dominated by the $P$ and $PE$-type diagrams,
while contributions from the $W$ exchange ones suffer from severe CKM suppression.
Furthermore, the contributions from $B$ and $PB$-type exchange diagrams are predicted to be vanishingly small.

After the standard PQCD calculations,  we obtain the factorization formulas for the invariant amplitudes,
which can be transformed into the helicity amplitudes for more convenient analysis of the asymmetry parameters in the angular distribution.
Angular momentum conservation allows four complex helicity amplitudes to contribute in the decay under scrutiny.
Differing from the calculations in GFA, where the helicity amplitudes share the same complex phase,
the strong phase in  our approach is primarily derived from nonfactorizable contribution that is special to each helicity amplitude.
Both the moduli and phases of these helicity amplitudes could be predicted in PQCD,
which allows one to compute the nonzero triple product asymmetries originating from the interference among various helicity amplitudes.
We found  that the negative helicity component of the $\Lambda$ baryon and longitudinally polarized $\phi$ meson in the final states are preferred.
This pattern observed is  in line with the expectation from the heavy-quark limit and the left-handed nature of the weak interaction and  the GFA prediction.

Our prediction of $6.9^{+1.9+1.8}_{-2.0-1.6}\times10^{-6}$  for its branching ratio is comparable with the world averages from PDG and HFLAV,
whereas the theoretical estimates based on GFA and QCDF  have shown sizeable deviations.
The representative theoretical uncertainties from the nonperturbative parameters in $\Lambda_b$ LCDAs  and the hard scale were taken into account,
which can reach $30\%$ in magnitude.
However, most of the asymmetry observables
are  less sensitive to the variations of hadronic parameters owing to the cancellations of uncertainties in the ratios.
The direct $CP$ asymmetry is estimated to reach the percent level,
which is consistent with the results from GFA and QCDF.
These obtained asymmetries can be confronted with the future data.

We give the first theoretical estimates of the triple product asymmetries in the polarized decay distributions,
which can be expressed by the imaginary part of bilinear combinations of the helicity amplitudes,
the asymmetry parameter $\alpha_\Lambda$ related to  $\Lambda\rightarrow p\pi$ decay, and the $\Lambda_b$ polarization fraction $P_b$.
For numerical estimations,
we have used the value of $\alpha_\Lambda$ as given in PDG and set $P_b=0.1$.
The predicted values of TPAs for $\Lambda_b\rightarrow \Lambda \phi$ and their $CP$ conjugate counterparts are in the range from $10^{-4}$ to $10^{-2}$.
Among these asymmetries, only $A^1_T$ is independent of $P_b$ and has a largest value at $10^{-2}$ level,
which can be measured with the unpolarized angular distribution.
Combining the TPAs in $\Lambda_b$ and $\bar{\Lambda}_b$ decays, we obtain the PQCD predictions on the true TPAs,
which are tiny, of order  $10^{-2}$ or even lower.
The smallness of these true asymmetries  are compatible with $CP$ conservation.
Hence, the measurement of a large true TPAs  would be
a clean indication of new $CP$ violation mechanism beyond the standard model.

\begin{acknowledgments}
We would like to acknowledge Yue-Long Shen and Fu-Sheng Yu for helpful discussions.
This work is supported by National Natural Science Foundation
of China under Grants No.12075086 and  No.11605060 and the Natural Science Foundation of Hebei Province
under Grants No.A2021209002 and  No.A2019209449.
\end{acknowledgments}

\begin{appendix}
\section{FACTORIZATION FORMULAS}\label{sec:for}
Following the conventions in Ref.~\cite{prd105073005}, we provide some details about the factorization formulas in Eq.~(\ref{eq:amp}).
The combinations of the Wilson coefficients $a_{R_{ij}}^{\sigma}$ are collected in Table~\ref{tab:wilson}.
The virtualities of the internal propagators $t_{A,B,C,D}$ and the expressions of $[\mathcal{D}b]$ and $\Omega_{R_{ij}}$ for the    exchange diagrams
are gathered in Tables~\ref{tab:ttt} and ~\ref{tab:bb}, respectively,
where the auxiliary functions $h_{1,2,3}$ and the Bessel function $K_0$ can be found in~\cite{prd105073005}.
The corresponding forms for the $P$-type diagrams can be found in Refs.~\cite{prd105073005,220604501} and shall not be repeated here.

In Table~\ref{tab:amppc}, we give the expressions of $H^{LL,LR,SP}_{R_{ij}}$ in the invariant amplitudes $A_1^{T,L}$ and $A_2^L$
for some main contributing Feynman diagrams,
while the remaining ones can be derived in a similar way.
The corresponding formulas for those $B$ terms can be obtained by the following replacement:
\begin{eqnarray}
B_1^{T,L}=A_1^{T,L}|_{r_\phi\rightarrow -r_\phi,r_\Lambda\rightarrow -r_\Lambda, \Phi^{T}\rightarrow -\Phi^{T}}, \quad
B_2^L=A_2^L|_{r_\phi\rightarrow -r_\phi,r_\Lambda\rightarrow -r_\Lambda,\Phi^{V}\rightarrow -\Phi^{V},\Phi^{A}\rightarrow -\Phi^{A}}.
\end{eqnarray}
\begin{table} [!htbh]
\footnotesize
	\caption{The expressions of $a^{LL}$, $a^{LR}$ and $a^{SP}$ in Eq.~(\ref{eq:amp}) for the exchange topological diagrams.} 
	\newcommand{\tabincell}[2]{\begin{tabular}{@{}#1@{}}#2\end{tabular}}
	\label{tab:wilson}
	\begin{tabular}[t]{lccc}
		\hline\hline
		$R_{ij}$        &$a^{LL}$      &$a^{LR}$     &$a^{SP}$             \\ \hline
		$P_{a1,a2,a3,a5,b1,b2,b4}$       &$V_{tb}V_{ts}^*[\frac{4}{3}(C_3+C_4)-\frac{2}{3}(C_9+C_{10})]$    &$V_{tb}V_{ts}^*[C_6+\frac{1}{3}C_5-\frac{1}{2}C_8-\frac{1}{6}C_7]$     &$V_{tb}V_{ts}^*[C_5+\frac{1}{3}C_6-\frac{1}{2}C_7-\frac{1}{6}C_8]$     \\				
		$P_{a6,a7,b6,b7,c1,c2,d1,d2}$    &$V_{tb}V_{ts}^*[\frac{1}{3}(C_3+C_4)-\frac{1}{6}(C_9+C_{10})]$   &$V_{tb}V_{ts}^*[\frac{1}{3}C_5-\frac{1}{6}C_7]$       &$V_{tb}V_{ts}^*[\frac{1}{3}C_6-\frac{1}{6}C_8]$                 \\
		$P_{c5,c7,d6}$      &$V_{tb}V_{ts}^*[\frac{1}{12}(C_3+C_4)-\frac{1}{24}(C_9+C_{10})]$                &$V_{tb}V_{ts}^*[\frac{1}{3}C_5-\frac{1}{4}C_6-\frac{1}{6}C_7+\frac{1}{8}C_8]$&$V_{tb}V_{ts}^*[\frac{1}{3}C_6-\frac{1}{4}C_5-\frac{1}{6}C_8+\frac{1}{8}C_7]$\\
						
		$PE_{a1-a7,e1-e4,f4}$ &$V_{tb}V_{ts}^*[\frac{2}{3}(C_3-C_4)+\frac{1}{6}(C_9-C_{10})]$     &$V_{tb}V_{ts}^*[\frac{2}{3}(C_5-C_6)+\frac{1}{6}(C_7-C_8)]$   &$\cdots$\\		
				
		$PE_{b1,b3,b5,b7}$ &$\frac{1}{3}V_{tb}V_{ts}^*[2C_3+4C_4+\frac{1}{2}C_9+C_{10}]$     &$\frac{1}{3}V_{tb}V_{ts}^*[2C_5+4C_6+\frac{1}{2}C_7+C_8]$
		& $\cdots$\\			
			
		$PE_{b2,b4,b6}$ &$\frac{1}{3}V_{tb}V_{ts}^*[2C_3-\frac{1}{2}C_4+\frac{1}{2}C_9-\frac{1}{8}C_{10}]$              &$\frac{1}{3}V_{tb}V_{ts}^*[2C_5-\frac{1}{2}C_6+\frac{1}{2}C_7-\frac{1}{8}C_8]$  & $\cdots$\\		
				
		$PE_{c1,c4,c5,c7}$ &$-\frac{1}{3}V_{tb}V_{ts}^*[4C_3+2C_4+C_9+\frac{1}{2}C_{10}]$    &$-\frac{1}{3}V_{tb}V_{ts}^*[4C_5+2C_6+C_7+\frac{1}{2}C_8]$
		&  $\cdots$  \\		
				
		$PE_{c2,c3,c6}$ &$\frac{1}{3}V_{tb}V_{ts}^*[\frac{1}{2}C_3-2C_4+\frac{1}{8}C_9-\frac{1}{2}C_{10}]$         &$\frac{1}{3}V_{tb}V_{ts}^*[\frac{1}{2}C_5-2C_6+\frac{1}{8}C_7-\frac{1}{2}C_8]$  &  $\cdots$ \\		
				
		$PE_{d1,d2,d4,d5,d7}$ &$\frac{2}{3}V_{tb}V_{ts}^*[-2C_3+2C_4-\frac{1}{2}C_9+\frac{1}{2}C_{10}]$           &$\frac{2}{3}V_{tb}V_{ts}^*[-2C_5+2C_6-\frac{1}{2}C_7+\frac{1}{2}C_8]$
		& $\cdots$\\		
				
		$PE_{d3,d6}$ &$\frac{1}{12}V_{tb}V_{ts}^*[2C_3-2C_4+\frac{1}{2}C_9-\frac{1}{2}C_{10}]$            &$\frac{1}{12}V_{tb}V_{ts}^*[2C_5-2C_6+\frac{1}{2}C_7-\frac{1}{2}C_8]$ & $\cdots$ \\	
			
		$PE_{f1,f2}$ &$-\frac{1}{3}V_{tb}V_{ts}^*[\frac{5}{2}C_3+2C_4+\frac{5}{8}C_9+\frac{1}{2}C_{10}]$        &$-\frac{1}{3}V_{tb}V_{ts}^*[\frac{5}{2}C_5+2C_6+\frac{5}{8}C_7+\frac{1}{2}C_8]$ & $\cdots$ \\	
			
		$PE_{f3}$ &$\frac{1}{3}V_{tb}V_{ts}^*[2C_3+\frac{5}{2}C_4+\frac{1}{2}C_9+\frac{5}{8}C_{10}]$            &$\frac{1}{3}V_{tb}V_{ts}^*[2C_5+\frac{5}{2}C_6+\frac{1}{2}C_7+\frac{5}{8}C_8]$ & $\cdots$ \\	
			
		$PE_{g1}$ &$-\frac{3}{4}V_{tb}V_{ts}^*[2C_3+\frac{1}{2}C_9]$          &$-\frac{3}{4}V_{tb}V_{ts}^*[2C_5+\frac{1}{2}C_7]$   & $\cdots$ \\
		
		$PE_{g2}$ &$\frac{3}{4}V_{tb}V_{ts}^*[-2C_3+2C_4-\frac{1}{2}C_9+\frac{1}{2}C_{10}]$          &$\frac{3}{4}V_{tb}V_{ts}^*[-2C_5+2C_6-\frac{1}{2}C_7+\frac{1}{2}C_8]$   & $\cdots$\\	
			
		$PE_{g3}$ &$\frac{3}{4}V_{tb}V_{ts}^*[2C_4+\frac{1}{2}C_{10}]$          &$\frac{3}{4}V_{tb}V_{ts}^*[2C_6+\frac{1}{2}C_8]$   & $\cdots$\\
		
        $PE_{g4}$ &$0$          &$0$   & $\cdots$\\
		
		$E_{a1-a7,e1-e4,f4}$ &$\frac{1}{3}V_{ub}V_{us}^*[C_1-C_2]$  &$\cdots$ &$\cdots$  \\						
		$E_{b1,b3,b5,b7}$ &$\frac{1}{3}V_{ub}V_{us}^*[C_1+2C_2]$          &$\cdots$& $\cdots$ \\				
		$E_{b2,b4,b6}$ &$\frac{1}{3}V_{ub}V_{us}^*[C_1-\frac{1}{4}C_2]$    &$\cdots$& $\cdots$\\				
		$E_{c1,c4,c5,c7}$ &$-\frac{1}{3}V_{ub}V_{us}^*[2C_1+C_2]$          &$\cdots$& $\cdots$  \\				
		$E_{c2,c3,c6}$ &$\frac{1}{3}V_{ub}V_{us}^*[\frac{1}{4}C_1-C_2]$ &$\cdots$& $\cdots$ \\				
		$E_{d1,d2,d4,d5,d7}$ &$\frac{2}{3}V_{ub}V_{us}^*[-C_1+C_2]$   &$\cdots$& $\cdots$\\				
		$E_{d3,d6}$ &$\frac{1}{12}V_{ub}V_{us}^*[C_1-C_2]$           &$\cdots$& $\cdots$\\		
		$E_{f1,f2}$ &$-\frac{1}{3}V_{ub}V_{us}^*[\frac{5}{4}C_1-C_2]$       &$\cdots$& $\cdots$ \\		
		$E_{f3}$ &$\frac{1}{3}V_{ub}V_{us}^*[C_1+\frac{5}{4}C_2]$            &$\cdots$& $\cdots$ \\		
		$E_{g1}$ &$-\frac{3}{4}V_{ub}V_{us}^*C_1$          &$\cdots$& $\cdots$ \\
		$E_{g2}$ &$\frac{3}{4}V_{ub}V_{us}^*[-C_1+C_2]$          &$\cdots$& $\cdots$\\		
		$E_{g3}$ &$\frac{3}{4}V_{ub}V_{us}^*C_2$         &$\cdots$& $\cdots$\\
        $E_{g4}$ &$0$       &$\cdots$& $\cdots$\\
		$ B_{a1-a4,b1-b4}$   & $\frac{1}{4}V_{ub}V_{us}^*[-C_1+C_2]$ &$\cdots$& $\cdots$\\
        $ PB_{a1-a4,b1-b4}$   & $\frac{1}{4}V_{tb}V_{ts}^*[-2C_3+2C_4-\frac{1}{2}C_9+\frac{1}{2}C_{10}]$ &$\cdots$ & $\frac{1}{4}V_{tb}V_{ts}^*[-2C_5+2C_6-\frac{1}{2}C_7+\frac{1}{2}C_8]$\\    	
		\hline\hline
	\end{tabular}
\end{table}

\begin{table}[H]
\footnotesize
\centering
	\caption{The virtualities of the internal gluon $t_{A,B}$ and quark $t_{C,D}$ for the exchange topological diagrams.}
	\newcommand{\tabincell}[2]{\begin{tabular}{@{}#1@{}}#2\end{tabular}}
	\label{tab:ttt}
	\begin{tabular}[t]{lcll}
		\hline\hline
		&     &$R_{ij}$	      & \\\hline
		$\frac{t_A}{M^2}$
		
		&&$E_{a1-d7,}PE_{a1-d7,}B_{a1-b4,}PB_{a1-b4}$   &$x_3x_3'$      \\
		&&$E_{e1-e4,f1-f4,}PE_{e1-e4,f1-f4}$       &$(1-x_2') (x_3+y-1)$ \\
		&&$E_{g1-g4,}PE_{g1-g4}$                &$ x_1'(y-1) $ \\ \hline	
		
		$\frac{t_B}{M^2}$    &&\tabincell{l}{$E_{a1,a2,a5-a7,b1,b2,b5-b7,c1-c3,c6,c7,d1-d3,d6,d7,e1-e4,f1-f4,}$\\$PE_{a1-a5,b1-b5,c1-c5,d1-d5,e1-e4,f1-f4}$}     &$x_1'(y-1)$   \\	
		&&$E_{a3,a4,b3,b4,c4,c5,d4,d5,g1-g4,}PE_{a6,a7,b6,b7,c6,c7,d6,d7,g1-g4}$         &$ (1-x_2') (x_3+y-1)$ \\
		&&$B_{a1-a4,}PB_{a1-a4}$     &$x_3' (x_3+y-1)$  \\	
		&&$B_{b1-b4,}PB_{b1-b4}$     &$ x_3'(x_3-y)$ \\   \hline
		
		$\frac{t_C}{M^2}$    &&$E_{a1,b1,c1,d1,}PE_{a3,b3,c4,d4}$     &$x_3'(1-x_1) $   \\	
		&&$E_{a2,b2,c3,d3,}PE_{a2,b2,c3,d3}$     &$ x_3'(x_3-y)$  \\
		&&$E_{a3,b3,c2,c4,d4,}PE_{a6,b4,b6,c4,d6}$  &$ x_3' (x_3+y-1)$  \\
		&&$E_{a4,b4,c5,d5,}PE_{a7,b7,c7,d7}$     &$ x_3(1-x_2')$  \\
		&&$E_{a5,b5,c6,d6,}PE_{a1,b1,c1,d1}$     &$ x_3'(1-x_2)+x_2$  \\
		&&$E_{a6,e4,f4,}PE_{a4,e4,f4}$        &$x_3+y-1$   \\
		&&$E_{a7,b7,c7,d7,}PE_{a5,b5,c5,d5}$     &$x_3(1-x_1')$   \\
		&&$E_{b6,}PE_{c2}$              &$x_1'(x_1+y-1)-x_1-y+2$  \\		
		&&$E_{d2,e2,f2,}PE_{d2,e2,f2}$        &$(x_3-1) (1-x_2')$  \\
		&&$E_{e1,f1,}PE_{e3,f3}$           &$ (x_1-y)(x_2'-1)$  \\		
		&&$E_{e3,f3,}PE_{e1,f1}$           &$ x_2'(x_2-y)+1$  \\
		&&$E_{g1-g4,}PE_{g1-g4}$        &$ x_3x_3' $ \\
		&&$B_{a1-a4,b1-b4,}PB_{a1-a4,b1-b4}$     &$x_3' (x_3-1) $   \\	\hline	
		
		$\frac{t_D}{M^2}$       &&$E_{a1,a2,a5,a6,}PE_{a1-a4}$     &$(x_3'-1)(1-y)$   \\	
		&&$E_{a3,a4,a7,}PE_{a5-a7}$          &$x_3+y-1$  \\
		&&$E_{b1,b2,b7,}PE_{b3-b5}$          &$x_1'(x_1+y-1)-x_1-y+2$  \\
		&&$E_{b3-b6,g3,}PE_{c1,c2,c6,c7,g1}$     &$ x_2'(x_2-y)+1$  \\
		&&$E_{c1,c2,c4,c5,g1,}PE_{b3,b4,b6,b7,g3}$     &$ (x_1-y)(x_2'-1)$  \\
		&&$E_{c3,c6,c7,}PE_{b1,b2,b5}$          &$ x_1' (x_2+y-1)$  \\
		&&$E_{d1,d2,d6,d7,}PE_{d1,d2,d4,d5}$        &$-x_1'$  \\
		&&$E_{d3-d5,g2,}PE_{d3,d6,d7,g2,}B_{a4,b4,}PB_{a2,b2}$     &$(x_3-1) (1-x_2')$   \\
		&&$E_{e1-e4,}PE_{e1-e4}$        &$  (1-x_2')(y-1)$  \\
		&&$E_{f1-f4,}PE_{f1-f4}$        &$ x_1'(x_3+y-1)$  \\
		&&$E_{g4,}PE_{g4}$              &$ x_3+y$ \\
		&&$B_{a1,b1,}PB_{a1,b1}$           &$x_2(1-x_3')+1$  \\	
		&&$B_{a2,b2,}PB_{a4,b4}$           &$(x_3-1)(1-x_1')$  \\
		&&$B_{a3,b3,}PB_{a3,b3}$           &$ -x_1x_3'$  \\                  			
		\hline\hline
	\end{tabular}
\end{table}

\begin{table}[H]
	\footnotesize
	\caption{The expressions of $[\mathcal{D}b]$ and $\Omega_{R_{ij}}$ for the exchange topological diagrams.}
	\label{tab:bb}
	\begin{tabular}[t]{lccc}
		\hline\hline
		$R_{ij}$ & $[\mathcal{D}b]$ &$\Omega_{R_{ij}}$\\ \hline
		$E_{a1},PE_{a3}$ & $\int d^2 \textbf{b}_2d^2\textbf{b}_3d^2\textbf{b}'_2d^2\textbf{b}'_3$
		&$\frac{1}{(2\pi)^4}K_0(\sqrt{t_A}|\textbf{b}_2-\textbf{b}_3|)K_0(\sqrt{t_B}|\textbf{b}'_2|)K_0(\sqrt{t_C}|\textbf{b}_2|)K_0(\sqrt{t_D}|\textbf{b}'_2-\textbf{b}'_3-\textbf{b}_3|)$ \\
		
		$E_{a2},PE_{a2}$ & $\int d^2 \textbf{b}_qd^2\textbf{b}_3d^2\textbf{b}'_2d^2\textbf{b}'_3$
		&$\frac{1}{(2\pi)^4}K_0(\sqrt{t_A}|\textbf{b}_q-\textbf{b}'_3|)K_0(\sqrt{t_B}|\textbf{b}'_2|)K_0(\sqrt{t_C}|\textbf{b}_q-\textbf{b}_3-\textbf{b}'_3|)K_0(\sqrt{t_D}|\textbf{b}'_2-\textbf{b}'_3-\textbf{b}_3|)$ \\
		
		$E_{a3},PE_{a6}$ & $\int d^2 \textbf{b}_qd^2\textbf{b}_3d^2\textbf{b}'_2d^2\textbf{b}'_3$
		&$\frac{1}{(2\pi)^4}K_0(\sqrt{t_A}|\textbf{b}_q-\textbf{b}_3|)K_0(\sqrt{t_B}|\textbf{b}'_2|)K_0(\sqrt{t_C}|\textbf{b}_q-\textbf{b}_3-\textbf{b}'_3|)K_0(\sqrt{t_D}|\textbf{b}'_2-\textbf{b}'_3-\textbf{b}_3|)$ \\
		
		$E_{a4},PE_{a7}$ & $\int d^2 \textbf{b}_qd^2\textbf{b}_3d^2\textbf{b}'_2d^2\textbf{b}'_3$
		&$\frac{1}{(2\pi)^4}K_0(\sqrt{t_A}|\textbf{b}_3-\textbf{b}'_2|)K_0(\sqrt{t_B}|\textbf{b}_3-\textbf{b}_q-\textbf{b}'_2+\textbf{b}'_3|)K_0(\sqrt{t_C}|\textbf{b}_q-\textbf{b}'_3-\textbf{b}_3|)K_0(\sqrt{t_D}|\textbf{b}'_2-\textbf{b}'_3-\textbf{b}_3|)$ \\
		
		$E_{a5},PE_{a1}$ & $\int d^2 \textbf{b}_2d^2\textbf{b}_3d^2\textbf{b}'_2d^2\textbf{b}'_3$
		&$\frac{1}{(2\pi)^4}K_0(\sqrt{t_A}|\textbf{b}_3|)K_0(\sqrt{t_B}|\textbf{b}'_2|)K_0(\sqrt{t_C}|\textbf{b}_2|)K_0(\sqrt{t_D}|\textbf{b}_2+\textbf{b}'_2-\textbf{b}'_3-\textbf{b}_3|)$ \\
		
		$E_{a6},PE_{a4}$ & $\int d^2 \textbf{b}_3 d^2\textbf{b}'_2 d^2\textbf{b}'_3$
		&$K_0(\sqrt{t_B}|\textbf{b}'_2|)h_2(\textbf{b}'_2-\textbf{b}'_3,\textbf{b}'_2-\textbf{b}'_3-\textbf{b}_3,t_A,t_C,t_D)$ \\
		
		$E_{a7},PE_{a5}$ & $\int d^2 \textbf{b}_3 d^2\textbf{b}'_2 d^2\textbf{b}'_3$
		&$K_0(\sqrt{t_A}|\textbf{b}'_2-\textbf{b}'_3|)h_2(\textbf{b}'_2-\textbf{b}'_3-\textbf{b}_3,-\textbf{b}_3-\textbf{b}'_3,t_B,t_C,t_D)$ \\
		
		$E_{b1},PE_{c4}$ & $\int d^2 \textbf{b}_2d^2\textbf{b}_3d^2\textbf{b}'_2d^2\textbf{b}'_3$
		&$\frac{1}{(2\pi)^4}K_0(\sqrt{t_A}|\textbf{b}_2-\textbf{b}_3|)K_0(\sqrt{t_B}|\textbf{b}_3+\textbf{b}'_3|)K_0(\sqrt{t_C}|\textbf{b}_2-\textbf{b}_3+\textbf{b}'_2-\textbf{b}'_3|)K_0(\sqrt{t_D}|\textbf{b}'_2-\textbf{b}'_3-\textbf{b}_3|)$ \\
		
		$E_{b2},PE_{c3}$ & $\int d^2 \textbf{b}_qd^2\textbf{b}_3d^2\textbf{b}'_2d^2\textbf{b}'_3$
		&$\frac{1}{(2\pi)^4}K_0(\sqrt{t_A}|\textbf{b}_q-\textbf{b}'_3|)K_0(\sqrt{t_B}|\textbf{b}_3+\textbf{b}'_3|)K_0(\sqrt{t_C}|\textbf{b}_q-\textbf{b}'_2|)K_0(\sqrt{t_D}|\textbf{b}'_2-\textbf{b}'_3-\textbf{b}_3|)$ \\
		
		$E_{b3},PE_{c6}$ & $\int d^2 \textbf{b}_qd^2\textbf{b}_3d^2\textbf{b}'_2d^2\textbf{b}'_3$
		&$\frac{1}{(2\pi)^4}K_0(\sqrt{t_A}|\textbf{b}'_2-\textbf{b}'_3-\textbf{b}_q|)K_0(\sqrt{t_B}|\textbf{b}_3+\textbf{b}'_3|)K_0(\sqrt{t_C}|\textbf{b}_q-\textbf{b}'_2|)K_0(\sqrt{t_D}|\textbf{b}'_2-\textbf{b}'_3-\textbf{b}_3|)$ \\
		
		$E_{b4},PE_{c7}$ & $\int d^2 \textbf{b}_qd^2\textbf{b}_3d^2\textbf{b}'_2d^2\textbf{b}'_3$
		&$\frac{1}{(2\pi)^4}K_0(\sqrt{t_A}|\textbf{b}'_3|)K_0(\sqrt{t_B}|\textbf{b}'_2-\textbf{b}'_3-\textbf{b}_3-\textbf{b}_q|)K_0(\sqrt{t_C}|\textbf{b}_q-\textbf{b}'_2|)K_0(\sqrt{t_D}|\textbf{b}'_2-\textbf{b}'_3-\textbf{b}_3|)$ \\
		
		$E_{b5},PE_{c1}$ & $\int d^2 \textbf{b}_3 d^2\textbf{b}'_2 d^2\textbf{b}'_3$
		&$K_0(\sqrt{t_A}|\textbf{b}_3|)h_2(\textbf{b}'_2-\textbf{b}'_3-\textbf{b}_3,\textbf{b}'_2,t_B,t_C,t_D)$ \\
		
		$E_{b6},PE_{c2}$ & $\int d^2 \textbf{b}_3 d^2\textbf{b}'_2 d^2\textbf{b}'_3$
		&$K_0(\sqrt{t_B}|\textbf{b}_3+\textbf{b}'_3|)h_2(\textbf{b}'_2-\textbf{b}'_3,\textbf{b}'_2-\textbf{b}'_3-\textbf{b}_3,t_A,t_C,t_D)$ \\
		
		$E_{b7},PE_{c5}$ & $\int d^2 \textbf{b}_2d^2\textbf{b}_3d^2\textbf{b}'_2d^2\textbf{b}'_3$
		&$\frac{1}{(2\pi)^4}K_0(\sqrt{t_A}|\textbf{b}'_2-\textbf{b}'_3|)K_0(\sqrt{t_B}|\textbf{b}_3+\textbf{b}'_3|)K_0(\sqrt{t_C}|\textbf{b}_2-\textbf{b}_3+\textbf{b}'_2-\textbf{b}'_3|)K_0(\sqrt{t_D}|\textbf{b}_2|)$ \\
		
		$E_{c1},PE_{b3}$ & $\int d^2 \textbf{b}'_3 d^2\textbf{b}_2 d^2\textbf{b}_3$
		&$K_0(\sqrt{t_A}|\textbf{b}_2-\textbf{b}_3|)h_2(\textbf{b}_2,\textbf{b}_3+\textbf{b}'_3,t_B,t_C,t_D)$ \\
		
		$E_{c2},PE_{b4}$ & $\int d^2 \textbf{b}'_3 d^2\textbf{b}_2 d^2\textbf{b}_3$
		&$K_0(\sqrt{t_A}|\textbf{b}_2-\textbf{b}_3-\textbf{b}'_3|)h_2(\textbf{b}_3,\textbf{b}_2,t_B,t_C,t_D)$ \\
		
		$E_{c3},PE_{b2}$ & $\int d^2 \textbf{b}_qd^2\textbf{b}'_3d^2\textbf{b}_2d^2\textbf{b}_3$
		&$\frac{1}{(2\pi)^4}K_0(\sqrt{t_A}|\textbf{b}_q-\textbf{b}'_3|)K_0(\sqrt{t_B}|\textbf{b}_2-\textbf{b}_3-\textbf{b}'_3|)K_0(\sqrt{t_C}|\textbf{b}_q-\textbf{b}_3-\textbf{b}'_3|)K_0(\sqrt{t_D}|\textbf{b}_2|)$ \\
		
		$E_{c4},PE_{b6}$ & $\int d^2 \textbf{b}_qd^2\textbf{b}'_3d^2\textbf{b}_2d^2\textbf{b}_3$
		&$\frac{1}{(2\pi)^4}K_0(\sqrt{t_A}|\textbf{b}_q-\textbf{b}_3|)K_0(\sqrt{t_B}|\textbf{b}_2-\textbf{b}_3-\textbf{b}'_3|)K_0(\sqrt{t_C}|\textbf{b}_q-\textbf{b}_3-\textbf{b}'_3|)K_0(\sqrt{t_D}|\textbf{b}_2)$ \\
		
		$E_{c5},PE_{b7}$ & $\int d^2 \textbf{b}_qd^2\textbf{b}'_3d^2\textbf{b}_2d^2\textbf{b}_3$
		&$\frac{1}{(2\pi)^4}K_0(\sqrt{t_A}|\textbf{b}'_3|)K_0(\sqrt{t_B}|\textbf{b}_2-\textbf{b}_q|)K_0(\sqrt{t_C}|\textbf{b}_q-\textbf{b}_3-\textbf{b}'_3|)K_0(\sqrt{t_D}|\textbf{b}_2|)$ \\
		
		$E_{c6},PE_{b1}$ & $\int d^2 \textbf{b}_qd^2\textbf{b}'_3d^2\textbf{b}_2d^2\textbf{b}_3$
		&$\frac{1}{(2\pi)^4}K_0(\sqrt{t_A}|\textbf{b}'_3|)K_0(\sqrt{t_B}|\textbf{b}_2-\textbf{b}_3-\textbf{b}'_3|)K_0(\sqrt{t_C}|\textbf{b}_q-\textbf{b}_3-\textbf{b}'_3|)K_0(\sqrt{t_D}|\textbf{b}_2+\textbf{b}_q-\textbf{b}_3-\textbf{b}'_3|)$ \\
		
		$E_{c7},PE_{b5}$ & $\int d^2 \textbf{b}_2d^2\textbf{b}_3d^2\textbf{b}'_2d^2\textbf{b}'_3$
		&$\frac{1}{(2\pi)^4}K_0(\sqrt{t_A}|\textbf{b}'_2-\textbf{b}'_3|)K_0(\sqrt{t_B}|\textbf{b}_2-\textbf{b}_3-\textbf{b}'_3|)K_0(\sqrt{t_C}|\textbf{b}'_2-\textbf{b}'_3+\textbf{b}_3|)K_0(\sqrt{t_D}|\textbf{b}_2|)$ \\
		
		$E_{d1},PE_{d4}$ & $\int d^2 \textbf{b}_2d^2\textbf{b}_3d^2\textbf{b}'_3d^2\textbf{b}_q$
		&$\frac{1}{(2\pi)^4}K_0(\sqrt{t_A}|\textbf{b}_2-\textbf{b}_3|)K_0(\sqrt{t_B}|\textbf{b}_q|)K_0(\sqrt{t_C}|\textbf{b}_2|)K_0(\sqrt{t_D}|\textbf{b}_q-\textbf{b}_3-\textbf{b}'_3)$ \\
		
		$E_{d2},PE_{d2}$ & $\int   d^2\textbf{b}_3 d^2\textbf{b}'_3d^2\textbf{b}_q$
		&$K_0(\sqrt{t_B}|\textbf{b}_q|)h_2(\textbf{b}_q-\textbf{b}'_3,\textbf{b}_q-\textbf{b}_3-\textbf{b}'_3,t_A,t_C,t_D)$ \\
		
		$E_{d3},PE_{d3}$ & $\int   d^2\textbf{b}_3 d^2\textbf{b}'_3d^2\textbf{b}_q$
		&$K_0(\sqrt{t_A}|\textbf{b}_q-\textbf{b}'_3|)h_2(\textbf{b}_3+\textbf{b}'_3,\textbf{b}_q-\textbf{b}_3-\textbf{b}'_3,t_B,t_C,t_D)$ \\
		
		$E_{d4},PE_{d6}$ & $\int   d^2\textbf{b}_3 d^2\textbf{b}'_3d^2\textbf{b}_q$
		&$h_3(\textbf{b}_3-\textbf{b}_q,\textbf{b}_3+\textbf{b}'_3-\textbf{b}_q,-\textbf{b}_3+\textbf{b}'_3,t_A,t_B,t_C,t_D)$ \\
		
		$E_{d5},PE_{d7}$ & $\int d^2 \textbf{b}_3 d^2\textbf{b}'_3 d^2\textbf{b}_q$
		&$K_0(\sqrt{t_A}|\textbf{b}'_3|)K_0(\sqrt{t_B}|\textbf{b}_q|)h_1(\textbf{b}_q-\textbf{b}_3-\textbf{b}'_3,t_C,t_D)$ \\
		
		$E_{d6},PE_{d1}$ & $\int d^2 \textbf{b}_2d^2\textbf{b}_3d^2\textbf{b}'_3d^2\textbf{b}_q$
		&$\frac{1}{(2\pi)^4}K_0(\sqrt{t_A}|\textbf{b}_3|)K_0(\sqrt{t_B}|\textbf{b}_q|)K_0(\sqrt{t_C}|\textbf{b}_2|)K_0(\sqrt{t_D}|\textbf{b}_q+\textbf{b}_2-\textbf{b}'_3-\textbf{b}_3|)$ \\
		
		$E_{d7},PE_{d5}$ & $\int d^2 \textbf{b}'_2d^2\textbf{b}'_3d^2\textbf{b}_3d^2\textbf{b}_q$
		&$\frac{1}{(2\pi)^4}K_0(\sqrt{t_A}|\textbf{b}'_2-\textbf{b}'_3|)K_0(\sqrt{t_B}|\textbf{b}_q|)K_0(\sqrt{t_C}|\textbf{b}'_2-\textbf{b}'_3-\textbf{b}_3|)K_0(\sqrt{t_D}|\textbf{b}_q-\textbf{b}_3-\textbf{b}'_3|)$ \\
		
		$E_{e1},PE_{e3}$ & $\int d^2 \textbf{b}_2d^2\textbf{b}_3d^2\textbf{b}'_3d^2\textbf{b}_q$
		&$\frac{1}{(2\pi)^4}K_0(\sqrt{t_A}|\textbf{b}_2-\textbf{b}_3|)K_0(\sqrt{t_B}|\textbf{b}'_3|)K_0(\sqrt{t_C}|\textbf{b}_2|)K_0(\sqrt{t_D}|\textbf{b}_q-\textbf{b}_3-\textbf{b}'_3|)$ \\
		
		$E_{e2},PE_{e2}$ & $\int d^2 \textbf{b}'_2d^2\textbf{b}'_3d^2\textbf{b}_3d^2\textbf{b}_q$
		&$\frac{1}{(2\pi)^4}K_0(\sqrt{t_A}|\textbf{b}_q+\textbf{b}_3-\textbf{b}'_2|)K_0(\sqrt{t_B}|\textbf{b}'_3|)K_0(\sqrt{t_C}|\textbf{b}_q-\textbf{b}'_2|)K_0(\sqrt{t_D}|\textbf{b}_3+\textbf{b}'_3-\textbf{b}'_2|)$ \\
		
		$E_{e3},PE_{e1}$ & $\int d^2 \textbf{b}'_2d^2\textbf{b}'_3d^2\textbf{b}_3d^2\textbf{b}_q$
		&$\frac{1}{(2\pi)^4}K_0(\sqrt{t_A}|\textbf{b}'_3|)K_0(\sqrt{t_B}|\textbf{b}_q-\textbf{b}'_2-\textbf{b}'_3|)K_0(\sqrt{t_C}|\textbf{b}_q-\textbf{b}'_2|)K_0(\sqrt{t_D}|\textbf{b}_3+\textbf{b}'_3-\textbf{b}_q|)$ \\
		
		$E_{e4},PE_{e4}$ & $\int d^2 \textbf{b}'_2d^2\textbf{b}'_3d^2\textbf{b}_3d^2\textbf{b}_q$
		&$\frac{1}{(2\pi)^4}K_0(\sqrt{t_A}|\textbf{b}_3+\textbf{b}'_2-\textbf{b}_q|)K_0(\sqrt{t_B}|\textbf{b}'_3|)K_0(\sqrt{t_C}|\textbf{b}_q-\textbf{b}'_2|)K_0(\sqrt{t_D}|\textbf{b}_3+\textbf{b}'_3-\textbf{b}_q|)$ \\
		
		$E_{f1},PE_{f3}$ & $\int d^2 \textbf{b}'_2d^2\textbf{b}'_3d^2\textbf{b}_2d^2\textbf{b}_3$
		&$\frac{1}{(2\pi)^4}K_0(\sqrt{t_A}|\textbf{b}'_2+\textbf{b}'_3-\textbf{b}_2|)K_0(\sqrt{t_B}|\textbf{b}_3-\textbf{b}'_2|)K_0(\sqrt{t_C}|\textbf{b}_2|)K_0(\sqrt{t_D}|\textbf{b}_3+\textbf{b}'_3-\textbf{b}'_2|)$ \\
		
		$E_{f2},PE_{f2}$ & $\int d^2 \textbf{b}'_2d^2\textbf{b}'_3d^2\textbf{b}_3d^2\textbf{b}_q$
		&$\frac{1}{(2\pi)^4}K_0(\sqrt{t_A}|\textbf{b}_q-\textbf{b}'_3|)K_0(\sqrt{t_B}|\textbf{b}_3-\textbf{b}'_2|)K_0(\sqrt{t_C}|\textbf{b}_q-\textbf{b}'_2|)K_0(\sqrt{t_D}|\textbf{b}_3+\textbf{b}'_3-\textbf{b}'_2|)$ \\
		
		$E_{f3},PE_{f1}$ & $\int d^2 \textbf{b}'_2d^2\textbf{b}'_3d^2\textbf{b}_2d^2\textbf{b}_3$
		&$\frac{1}{(2\pi)^4}K_0(\sqrt{t_A}|\textbf{b}_2+\textbf{b}'_2-\textbf{b}'_3|)K_0(\sqrt{t_B}|\textbf{b}_3-\textbf{b}_2-\textbf{b}'_2|)K_0(\sqrt{t_C}|\textbf{b}_2|)K_0(\sqrt{t_D}|\textbf{b}_2-\textbf{b}_3+\textbf{b}'_2-\textbf{b}'_3|)$ \\
		
		$E_{f4},PE_{f4}$ & $\int d^2 \textbf{b}'_2d^2\textbf{b}'_3d^2\textbf{b}_3d^2\textbf{b}_q$
		&$\frac{1}{(2\pi)^4}K_0(\sqrt{t_A}|\textbf{b}'_2-\textbf{b}'_3|)K_0(\sqrt{t_B}|\textbf{b}_3-\textbf{b}_q|)K_0(\sqrt{t_C}|\textbf{b}'_2-\textbf{b}_q|)K_0(\sqrt{t_D}|\textbf{b}_3+\textbf{b}'_3-\textbf{b}_q|)$ \\
		
		$E_{g1},PE_{g3}$ & $\int d^2 \textbf{b}'_2 d^2\textbf{b}_2 d^2\textbf{b}_3$
		&$K_0(\sqrt{t_D}|\textbf{b}_2|)h_2(\textbf{b}_2-\textbf{b}_3,\textbf{b}_2-\textbf{b}'_2,t_A,t_B,t_C)$ \\
		
		$E_{g2},PE_{g2}$ & $\int d^2 \textbf{b}_q d^2\textbf{b}'_2 d^2\textbf{b}_3$
		&$K_0(\sqrt{t_D}|\textbf{b}_q-\textbf{b}'_2|)h_2(\textbf{b}'_2-\textbf{b}_3-\textbf{b}_q,-\textbf{b}_q,t_A,t_B,t_C)$ \\
		
		$E_{g3},PE_{g1}$ & $\int d^2\textbf{b}'_2 d^2 \textbf{b}_2  d^2\textbf{b}_3$
		&$K_0(\sqrt{t_D}|\textbf{b}_2|)h_2(-\textbf{b}_3,-\textbf{b}_2-\textbf{b}'_2,t_A,t_B,t_C)$ \\
		
		$E_{g4},PE_{g4}$ & $\int d^2\textbf{b}_q d^2\textbf{b}_3d^2 \textbf{b}'_3  $
		&$K_0(\sqrt{t_D}|\textbf{b}_q-\textbf{b}'_3|)h_2(\textbf{b}_q-\textbf{b}_3-\textbf{b}'_3,\textbf{b}_3+\textbf{b}'_3,t_A,t_B,t_C)$ \\

        $B_{a1},PB_{a1}$ & $\int d^2 \textbf{b}_qd^2\textbf{b}_2d^2\textbf{b}_3$
		&$K_0(\sqrt{t_B}|\textbf{b}_q|)K_0(\sqrt{t_D}|\textbf{b}_2|)h_1(\textbf{b}_3-\textbf{b}_q,t_A,t_C)$ \\
		
		$B_{a2},PB_{a4}$ & $\int d^2 \textbf{b}_qd^2\textbf{b}'_2d^2\textbf{b}_3$
		&$K_0(\sqrt{t_B}|\textbf{b}_q|)K_0(\sqrt{t_D}|\textbf{b}'_2|)h_1(\textbf{b}'_2+\textbf{b}_3-\textbf{b}_q,t_A,t_C)$ \\
		
		$B_{a3},PB_{a3}$ & $\int d^2 \textbf{b}_qd^2\textbf{b}_2d^2\textbf{b}'_3$
		&$K_0(\sqrt{t_B}|\textbf{b}_q|)K_0(\sqrt{t_D}|\textbf{b}_2|)h_1(\textbf{b}'_3+\textbf{b}_2+\textbf{b}_q,t_A,t_C)$ \\
		
		$B_{a4},PB_{a2}$ & $\int d^2 \textbf{b}_qd^2\textbf{b}'_3d^2\textbf{b}_3$
		&$K_0(\sqrt{t_B}|\textbf{b}_q|)K_0(\sqrt{t_D}|\textbf{b}_3+\textbf{b}'_3|)h_1(\textbf{b}'_3+\textbf{b}_q,t_A,t_C)$\\
		
		$B_{b1},PB_{b1}$ & $\int d^2 \textbf{b}_qd^2\textbf{b}_2d^2\textbf{b}_3$
		&$K_0(\sqrt{t_B}|\textbf{b}_q|)K_0(\sqrt{t_D}|\textbf{b}_2|)h_1(\textbf{b}_3+\textbf{b}_q,t_A,t_C)$ \\
		
		$B_{b2},PB_{b4}$ & $\int d^2 \textbf{b}_qd^2\textbf{b}'_2d^2\textbf{b}_3$
		&$K_0(\sqrt{t_B}|\textbf{b}_q|)K_0(\sqrt{t_D}|\textbf{b}'_2|)h_1(\textbf{b}'_2+\textbf{b}_3+\textbf{b}_q,t_A,t_C)$\\
		
		$B_{b3},PB_{b3}$ & $\int d^2 \textbf{b}_qd^2\textbf{b}_2d^2\textbf{b}_3$
		&$K_0(\sqrt{t_B}|\textbf{b}_q|)K_0(\sqrt{t_D}|\textbf{b}_2|)h_1(\textbf{b}_3-\textbf{b}_2+\textbf{b}_q,t_A,t_C)$ \\
		
		$B_{b4},PB_{b2}$ & $\int d^2 \textbf{b}_qd^2\textbf{b}'_2d^2\textbf{b}_3$
		&$K_0(\sqrt{t_B}|\textbf{b}_q|)K_0(\sqrt{t_D}|\textbf{b}'_2|)(\textbf{b}_3-\textbf{b}'_2+\textbf{b}_q,t_A,t_C)$ \\
		
		\hline\hline
	\end{tabular}
\end{table}

\begin{table}[H]
	\footnotesize
	\centering
	\caption{The expressions of $H^{LL,LR,SP}_{R_{ij}}$ in the invariant amplitudes $A_1^{T,L}$ and $A_2^L$ for the diagrams $P_{c7,d6}$ and $PE_{b3}$.}
	\newcommand{\tabincell}[2]{\begin{tabular}{@{}#1@{}}#2\end{tabular}}
	\label{tab:amppc}
	\begin{tabular}[t]{lccc}
		\hline\hline
		$$  &$\frac{A_1^T}{16M^4}$&$\frac{A_1^L}{16M^4}$&$\frac{A_2^L}{16M^4}$\\ \hline
		$H_{P_{c7}}^{LL}$
		&\tabincell{c}{$r_{\phi} \Psi _3^{-+} (r_{\Lambda }-1) \Phi ^T (x_3+y-1)$\\$ (y-x_2) (\phi _V^a-\phi _V^v)$}
		&\tabincell{c}{$r_{\phi} \Psi _3 ^{-+} (r_{\Lambda }-1) \Phi ^T \phi _V $\\$(x_2-y)(x_3+y-1)$}
		&\tabincell{c}{$2 r_{\phi} \phi _V (\Phi ^A (-\Psi _4 x_2' x_3' r_{\Lambda }-2(x_3-1) $\\$y \Psi _2 (r_{\Lambda }+1)+2 x_2 \Psi _2(r_{\Lambda }+1)$\\$ (x_3+y-1)-2 y^2 \Psi _2 (r_{\Lambda}+1))+\Phi ^T$\\$ (\Psi _3 ^{-+} (x_3+y-1) (x_2'r_{\Lambda }-x_2+y)$\\$+x_3' \Psi _3 ^{+-} (r_{\Lambda }+1)(y-x_2))$\\$+\Psi _4 x_2' x_3' r_{\Lambda } \Phi ^V)$}\\\\
		$H_{P_{c7}}^{LR}$
		&\tabincell{c}{$ (r_{\Lambda }-1) \Phi ^T \phi _V^T (\Psi _3^{-+} x_2' (x_3+y-1)$\\$+x_3' \Psi _3^{+-}(y+x_2))$}
		&\tabincell{c}{$2 (r_{\Lambda }-1) \Phi ^T \phi _V^t (\Psi _3 ^{-+} x_2'(x_3+y-1)$\\$-y x_3' \Psi _3 ^{+-}+x_2 x_3' \Psi _3 ^{+-})$}
		&\tabincell{c}{$-r_{\phi}^2 \phi _V^s (r_{\Lambda } \Phi ^T (\Psi _3 ^{-+}+\Psi _3 ^{+-})(x_2' (x_3+y-1)$\\$+x_3' (y-x_2))-2 \Psi _4(x_2' (x_3+y-1) (\Phi ^A+\Phi ^V)$\\$+x_3'(x_2-y) (\Phi ^A-\Phi ^V)))-2 \Phi ^T \phi _V^t$\\$(\Psi _3 ^{-+} x_2' (x_3+y-1)+x_3' \Psi _3 ^{+-}(x_2-y))$}\\\\
		$H_{P_{c7}}^{SP}$
		&\tabincell{c}{$r_{\phi} \Psi _3^{+-} (r_{\Lambda }-1) \Phi ^T (y-x_2) $\\$(x_3+y-1) (\phi _V^a+\phi_V^v)$}
		&\tabincell{c}{$r_{\phi} \Psi _3 ^{-+} (r_{\Lambda }-1) \Phi ^T \phi _V $\\$(y-x_2)(x_3+y-1)$}
		&\tabincell{c}{$2 r_{\phi} \phi _V (\Phi ^A (-\Psi _4 x_2' x_3' r_{\Lambda }-2(x_3-1) y \Psi_2 $\\$(r_{\Lambda }+1)+2 x_2 \Psi _2(r_{\Lambda }+1) (x_3+y-1)$\\$-2 y^2 \Psi _2 (r_{\Lambda}+1))+\Phi ^T (\Psi _3 ^{-+} x_2' (r_{\Lambda }+1)$\\$(x_3+y-1)+\Psi _3 ^{+-} (x_2-y) (x_3-x_3'$\\$ r_{\Lambda}+y-1))-\Psi _4 x_2' x_3' r_{\Lambda } \Phi ^V)$}\\\\
		
		$H_{P_{d6}}^{LL}$
		&\tabincell{c}{$-r_{\phi} \Phi ^A (r_{\Lambda } (\Psi _4 (x_2' ((2 (1-x_2-y)-x_3) $\\$\phi _V^a+x_3\phi _V^v)+x_3' (1-x_2-y) (\phi _V^a+\phi _V^v))$\\$+2 \Psi _2 (x_2+y-1)(y-x_1) (\phi _V^a-\phi _V^v))$\\$+\Psi _4 (1-x_1')(x_2+y-1) (\phi _V^a+\phi _V^v))$\\$-\frac{1}{2} r_{\phi} (1-x_1') r_{\Lambda } \Phi^T (x_2+y-1) (\phi _V^a+\phi _V^v)$\\$ (\Psi _3^{-+}+\Psi _3^{+-})+r_{\phi} \Psi _4 \Phi ^V(x_2' \phi _V^a (x_3 r_{\Lambda }$\\$+x_2+y-1)-x_3' (r_{\Lambda }-1) (x_2+y-1)$\\$(\phi _V^a+\phi _V^v)+x_2' \phi _V^v (-r_{\Lambda } (2(x_2+y-1)$\\$+x_3)+x_2+y-1))$}
		&\tabincell{c}{$\frac{1}{2} r_{\phi} \phi _V (2 \Phi ^A (r_{\Lambda } (\Psi_4 x_3'(x_2+y-1)$\\$-x_3 \Psi _4 x_2'+2 \Psi_2 (x_2+y-1)(y-x_1))$\\$+\Psi_4 (1-x_1')(1-x_2-y))-(1-x_1')$\\$ r_{\Lambda } \Phi ^T(x_2+y-1) (\Psi_3 ^{-+}+\Psi _3 ^{+-})$\\$-2 \Psi _4 \Phi ^V(x_2' (r_{\Lambda } (2 x_2+x_3+2 y-2)$\\$-x_2-y+1)+x_3'(r_{\Lambda }-1) $\\$(x_2+y-1)))$}
		&\tabincell{c}{$r_{\phi} \phi _V (-2 \Phi ^A (\Psi _4 x_3' (x_2+y-1)-x_3 \Psi _4x_2'$\\$+2 \Psi _2 (x_2+y-1) (y-x_1))+\Phi ^T$\\$(2 (1-x_1') \Psi _3^{+-}(x_2+y-1)-r_{\Lambda }(\Psi _3 ^{-+} x_2'$\\$ (2 x_2+x_3+2 y-2)+x_3' (x_2+y-1)$\\$(\Psi _3 ^{-+}-\Psi_3 ^{+-})+x_3 x_2' \Psi _3 ^{+-}))+2 \Psi _4\Phi ^V$\\$ (x_2' (2 x_2+x_3+2 y-2)$\\$+x_3'(x_2+y-1)))$}\\\\
		$H_{P_{d6}}^{LR}$
		&$(x_1'-1)  (r_{\Lambda }-1)  (x_2+y-1)\Psi _3^{+-} \Phi ^T\phi _V^T$
		&$(x_1'-1)  (r_{\Lambda }-1) (x_2+y-1)\Psi _3 ^{+-} \Phi ^T\phi _V^t$
		&\tabincell{c}{$2 (1-x_1') \Psi_3 ^{+-} \Phi ^T (x_2+y-1) \phi _V^t$\\$-r_{\phi}^2 \phi_V^s (2 \Phi ^A (2 \Psi_2 (x_2+y-1)(y-x_1)$\\$-\Psi_4 (x_2' (2 x_2+x_3+2 y-2)$\\$+x_3'(x_2+y-1)))+r_{\Lambda } \Phi ^T $\\$(\Psi _3 ^{-+}+\Psi _3 ^{+-}) (x_3 x_2'-x_3' (x_2+y-1))$\\$-2 \Psi_4 \Phi ^V(x_3 x_2'-x_3' (x_2+y-1)))$}\\\\
		$H_{P_{d6}}^{SP}$
		&\tabincell{c}{$r_{\phi} \Psi_4 (\phi _V^a (x_3' (r_{\Lambda }-1) (y-x_1) (\Phi ^A-\Phi^V)$\\$+(x_1'-1) (\Phi ^A ((2 r_{\Lambda }-1) (y-x_1)-x_3 r_{\Lambda})$\\$+\Phi ^V (y-x_3 r_{\Lambda }-x_1)))+\phi _V^v (x_3' (r_{\Lambda }-1)$\\$(y-x_1) (\Phi ^V-\Phi ^A)+(x_1'-1) (\Phi ^A (y-x_3 r_{\Lambda}$\\$-x_1)+\Phi ^V ((2 r_{\Lambda }-1) (y-x_1)-x_3 r_{\Lambda}))))$\\$+2 r_{\phi} \Psi_2 \Phi ^A (r_{\Lambda }-1) (y-x_1)(x_2+y-1) $\\$(\phi _V^a-\phi _V^v)+\frac{1}{2} r_{\phi} \Phi ^T (-\Psi_3^{+-}r_{\Lambda} (\phi _V^a-\phi _V^v)$\\$ (x_3' (y-x_1)+(x_1'-1) (3y-3 x_1-2 x_3))$\\$-\Psi _3^{-+} (y-x_1) (\phi _V^a (r_{\Lambda } (2x_2-x_2'+2y-2)$\\$+2 (1-x_2-y))+\phi _V^v (2 (1-x_2-y)$\\$-r_{\Lambda } (2x_2-x_2'+2y-2))))$}
		&\tabincell{c}{$\frac{1}{2} r_{\phi} \phi _V (-2 \Phi ^A (\Psi _4 x_2' (x_3 (1-r_{\Lambda})+x_2$\\$+y-1)+\Psi _4 x_3' r_{\Lambda } (x_2+y-1)+2 \Psi _2$\\$(r_{\Lambda }-1) (x_2+y-1)(y-x_1))-\Phi ^T $\\$(\Psi _3 ^{-+} (y-x_1)(r_{\Lambda } (x_2'+2 x_2+2 y-2)$\\$-2(x_2+y-1))+\Psi _3 ^{+-} r_{\Lambda } (x_2' $\\$(3 x_2+x_3+3y-3)+2 x_3' (x_2+y-1)))$\\$-2 \Psi _4 \Phi ^V (x_2'(r_{\Lambda } (2 x_2+x_3+2 y-2)$\\$x_1-y)+x_3' r_{\Lambda} (x_2+y-1)))$}
		&\tabincell{c}{$2 r_{\phi} \phi _V (-\Phi ^A r_{\Lambda }(\Psi_4 x_3'(x_2+y-1)$\\$-x_3 \Psi _4 x_2'+2 \Psi _2 (x_2+y-1)(y-x_1))$\\$+\Psi_3 ^{-+} \Phi ^T (x_2+y-1)(y-x_1)$\\$-\Psi _4 r_{\Lambda } \Phi ^V (x_2' (2 x_2+x_3+2y-2)$\\$+x_3' (x_2+y-1)))$}\\\\

		$H_{PE_{b3}}^{LL}$
		&\tabincell{c}{$(1-x_2') ((1-x_1) \Phi ^T (\Psi_3^{-+}+\Psi_3^{+-})$\\$ (r_{\Lambda } (r_{\phi}\phi _V^a+r_{\phi} \phi _V^v-\phi _V^T)+\phi _V^T)$\\$-\Psi _4 \Phi ^A x_3' r_{\Lambda } \phi_V^T-\Psi _4 x_3' r_{\Lambda } \Phi ^V \phi _V^T)$}
		&\tabincell{c}{$(1-x_2') (\Psi _4 x_3' r_{\Lambda } (-\Phi ^A-\Phi ^V) \phi _V^t$\\$+(1-x_1) \Phi ^T(\Psi _3^{-+}+\Psi _3^{+-})$\\$ (r_{\phi} r_{\Lambda } \phi _V-(r_{\Lambda }-1) \phi _V^t))$}
		&\tabincell{c}{$x_3' (2 \Psi _4 ((x_2'-1) r_{\Lambda } (\Phi ^A+\Phi ^V) \phi _V^t-r_{\phi} \phi _V$\\$(\Phi ^A-\Phi ^V) (r_{\Lambda } (y+1-x_2'-x_1)-x_1+y)$\\$+r_{\phi}^2 r_{\Lambda }(y-x_1) (\Phi ^A+\Phi ^V) \phi _V^s)-r_{\phi} r_{\Lambda } \Phi ^T \phi _V$\\$ (y-x_1)(\Psi _3^{-+}+\Psi_3^{+-}))+(x_1-1) \Phi ^T $\\$(\Psi _3^{-+}+\Psi _3^{+-}) (2((x_2'-1) \phi _V^t+r_{\phi}^2 (y-x_1) \phi _V^s)$\\$-r_{\phi} r_{\Lambda } (r_{\phi} (x_2'+2x_1-2 y-1) \phi _V^s-2 (x_2'-1) \phi _V))$}\\\\
		$H_{PE_{b3}}^{LR}$
		&\tabincell{c}{$ (x_2'-1) x_3' r_{\Lambda } (\Phi ^A-\Phi ^V)\Psi _4 \phi _V^T$}
		&\tabincell{c}{$ (x_2'-1) x_3' r_{\Lambda } (\Phi ^A-\Phi ^V)\Psi _4 \phi _V^t$}
		&\tabincell{c}{$-r_{\phi}^2 \phi _V^s (x_3' (y-x_1) (2 \Psi _4 (r_{\Lambda }+1) (\Phi ^A-\Phi^V)$\\$+r_{\Lambda } \Phi ^T (\Psi _3^{-+}+\Psi _3^{+-}))+2 (x_1-1) (x_2'-1)$\\$r_{\Lambda } \Phi ^T (\Psi _3^{-+}+\Psi _3^{+-}))+2 \Psi _4 \Phi ^A (x_2'-1)$\\$ x_3'r_{\Lambda } (r_{\phi} \phi _V+\phi _V^t)+r_{\phi} (x_1-1) (x_2'-1)$\\$ (r_{\Lambda }+2)\Phi ^T \phi _V (\Psi _3^{-+}+\Psi _3^{+-})$\\$-2 \Psi _4 (x_2'-1) x_3' r_{\Lambda } \Phi ^V(\phi _V^t-r_{\phi} \phi _V)$}\\\\
		\hline\hline
	\end{tabular}
\end{table}

\end{appendix}

\end{document}